\documentclass[%
 reprint,
superscriptaddress,
 amsmath,amssymb,
 aps,
]{revtex4-2}

\usepackage{graphicx}
\usepackage{dcolumn}
\usepackage{bm}
\usepackage{cleveref}

\begin{document}

\preprint{APS/123-QED}

\title{Critical Disconnect Between Structural and Electronic Recovery in \\Amorphous GaAs during Recrystallization}

\author{Ellis Rae Kennedy}
\affiliation{Materials Physics and Applications Division, Los Alamos National Laboratory, Los Alamos, NM, USA}%
\thanks{LA-UR-25-30308}%

\author{Adric Jones}
\affiliation{Materials Science and Engineering, Arizona State University, Tempe, AZ, USA}%

\author{Yongqiang Wang}
\affiliation{Materials Science and Technology Division, Los Alamos National Laboratory, Los Alamos, NM, USA}%

\author{Miguel Pena}
\affiliation{Materials Science and Technology Division, Los Alamos National Laboratory, Los Alamos, NM, USA}%

\author{Hyosim Kim}
\affiliation{Materials Science and Technology Division, Los Alamos National Laboratory, Los Alamos, NM, USA}%

\author{Chengyu Song}
\affiliation{National Center for Electron Microscopy, Molecular Foundry, Lawrence Berkeley National Laboratory, Berkeley, CA, USA}%

\author{Farida Selim}
\affiliation{Materials Science and Engineering, Arizona State University, Tempe, AZ, USA}%

\author{Blas P. Uberuaga}
\affiliation{Materials Science and Technology Division, Los Alamos National Laboratory, Los Alamos, NM, USA}%

\author{Samuel Greer}
\affiliation{Physical Chemistry \& Applied Spectroscopy Division, Los Alamos National Laboratory, Los Alamos, NM, USA}%

\date{\today}

\begin{abstract}

Understanding the evolution of structure and functionality through amorphous to crystalline phase transitions is critical for predicting and designing devices for application in extreme conditions. Here, we consider both aspects of recrystallization of irradiated GaAs. We find that structural evolution occurs in two stages, a low temperature regime characterized by slow, epitaxial front propagation and a high-temperature regime above dominated by rapid growth and formation of dense nanotwin networks. We link aspects of this structural evolution to local ordering, or paracrystallinity, within the amorphous phase. Critically, the electronic recovery of the materials is not commensurate with this structural evolution. The electronic properties of the recrystallized material deviate further from the pristine material than do those of the amorphous phase, highlighting the incongruence between structural and electronic recovery and the contrasting impact of loss of long range order versus localized defects on the functionality of semiconducting materials.
\end{abstract}

\maketitle

\section{Introduction}

A material’s ability to recover its structure and functionality after disordering or amorphization is critical to long-term performance in both phase-change applications and harsh operating environments~\cite{srour1988radiation,johnston2013radiation,zhang2025unveiling,pearton2021review}. Amorphous materials are ubiquitous in modern technologies, spanning applications in microelectronics~\cite{jung2019amorphous,kamiya2010material}, catalysis~\cite{smith2014facile,zhang2021sodium}, and topological systems~\cite{ciocys2024establishing,zhang2023anomalous}. Yet, their atomic-scale structure remains difficult to resolve and connect directly to functional properties. Many amorphous solids exhibit short- and medium-range order, including paracrystalline motifs and non-random atomic packing, challenging the classical assumption of structural homogeneity~\cite{sorensen2020revealing,nishio2013universal,voyles2001structure}. Paracrystalline regions consist of locally correlated atomic configurations with variable spacing and orientation, representing an intermediate state between crystalline and fully disordered structures~\cite{nakayama2025symmetry}. Although such features are often invisible to conventional diffraction methods, they can play a decisive role in determining how materials evolve under external stimuli, particularly during phase transformations such as recrystallization. In semiconductors, this typically refers to the reordering of an amorphous phase back into a crystalline lattice. Debates surrounding amorphous silicon, for example, highlight how medium-range order and paracrystalline motifs challenge the traditional continuous random network model~\cite{nakayama2025symmetry,rosset2025signatures,treacy2012local}. The present work extends these discussions to amorphous GaAs, demonstrating that it retains structural memory capable of guiding its recrystallization dynamics.

Gallium arsenide (GaAs), a high-mobility semiconductor widely used in optoelectronics and photovoltaics~\cite{allen1960gallium}, provides an ideal platform to investigate the interplay between disorder, structural recovery, and functionality. GaAs performance is highly sensitive to defects such as twins and antisites, which degrade carrier mobility and modify the band structure~\cite{hamadani2022visualizing,williams1966determination}. Understanding how amorphous GaAs recrystallizes, particularly whether it preserves embedded local order that guides oriented regrowth, is essential for predicting and optimizing its performance following irradiation and thermal processing.

\begin{figure*}[t]
\centering
\includegraphics[width = 0.97\textwidth]{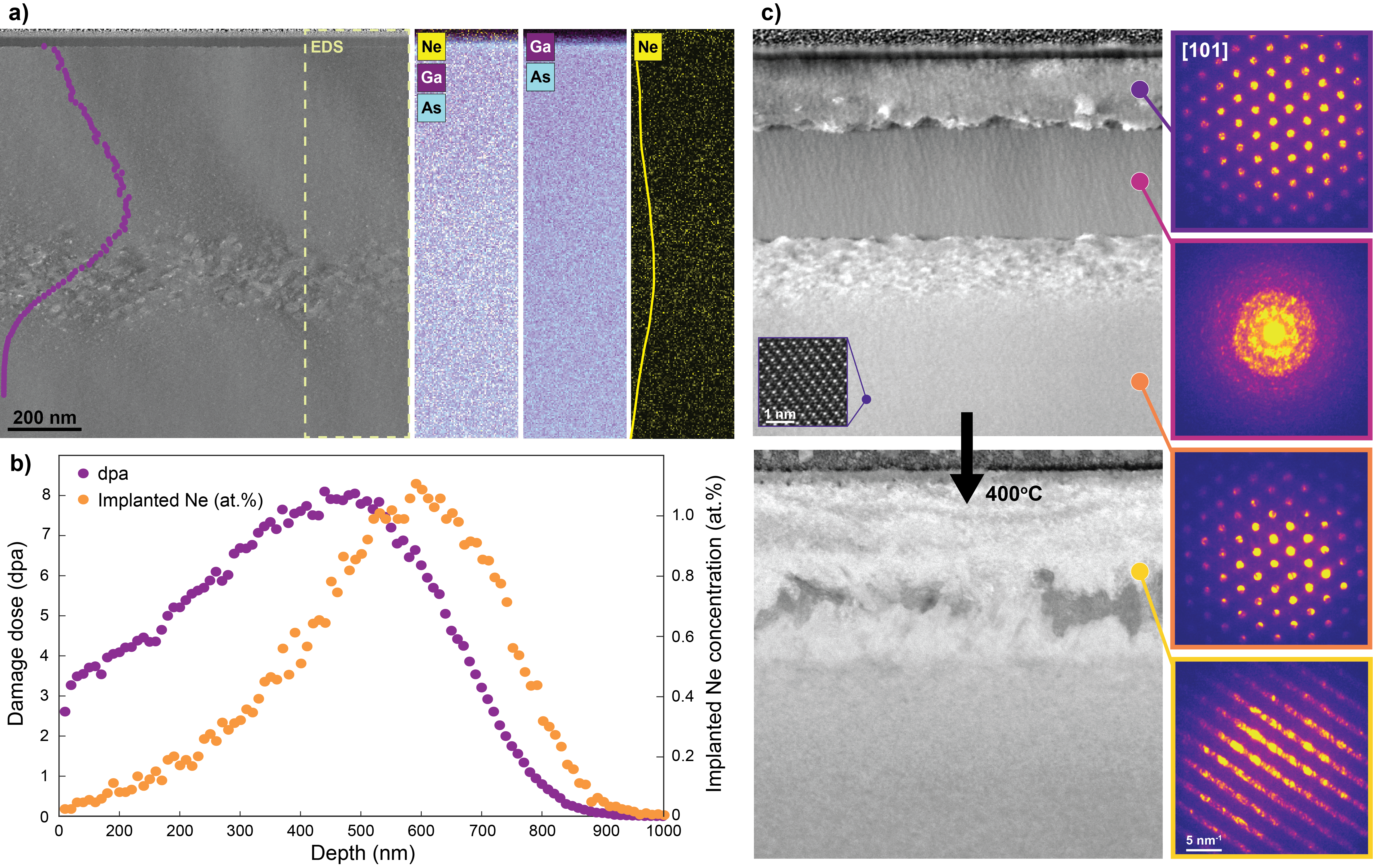}
  \caption{(a) Cross-sectional STEM image of [001]-oriented GaAs irradiated with 400 keV Ne$^{++}$ with a thin recrystallized layer at the top surface. A  approximately 510 nm thick amorphous layer extends below the recrystallized layer, underlain by a dense band of dislocation loops within the partially damaged crystal, and finally by undamaged pristine GaAs. EDS maps from the same region show uniform Ga and As distributions across both amorphous and crystalline zones. Ne is sparsely distributed, with a peak concentration of ~1 at.\% located near the dislocation loop zone. (b) SRIM simulations of displacement damage (dpa) and implanted Ne concentration as a function of depth. The dpa profile (replotted in (a)) shows a maximum of roughly 8 dpa at 480 nm, consistent with the depth of observed amorphization. (c) Over time, partial recrystallization occurs, primarily from the surface downward and to a lesser extent from the crystalline base upward. The recrystallized regions adopt the original [101] out-of-plane orientation of the cross-section of the undamaged GaAs. Inset: HRTEM image of undamamaged GaAs with resolved Ga and As atomic columns. After annealing at 400\textdegree{}C, the formerly amorphous layer becomes largely recrystallized, forming extensive twin domains aligned along the [111] directions and small polycrystalline pockets. Representative 4D-STEM diffraction patterns are shown (top to bottom) from the recrystallized upper layer, amorphous region, undamaged prsitine crystal, and twinned zones.}
  \label{fig:overview}
\end{figure*}

Here, we study the \textit{in situ} thermal recrystallization of GaAs irradiated with 400 keV Ne$^{++}$ ions along the [001] direction. This produces a structurally heterogeneous amorphous layer atop GaAs substrate. We identify two distinct recrystallization regimes: below 250\textdegree{}C, epitaxial regrowth preserves the original crystallographic orientation; above 250\textdegree{}C, dense \{111\} nanotwin networks emerge, modifying the local strain field and potentially altering electronic behavior. These observations point to complex interactions among interface morphology, thermal activation, and defect nucleation mechanisms during regrowth.

These insights are enabled by a symmetry- and strain-sensitive four-dimensional scanning transmission electron microscopy (4D-STEM) approach, which maps local structural variations at nanometer resolution~\cite{ophus2019four,savitzky2021py4dstem}. Even materials that appear fully amorphous under selected area electron diffraction (SAED) may harbor nanoscale heterogeneities, including paracrystalline or orientational domains~\cite{treacy2012local,kennedy2024exploring}, that can seed recrystallization. Detecting such structural memory, such as remnant symmetries and orientations inherited from its pre-irradiation state, in amorphous GaAs is inherently challenging and requires techniques capable of probing short- and medium-range order~\cite{gammer2018local,kennedy2025mapping}.

\begin{figure*}[t]
\centering
\includegraphics[width = 0.96\textwidth]{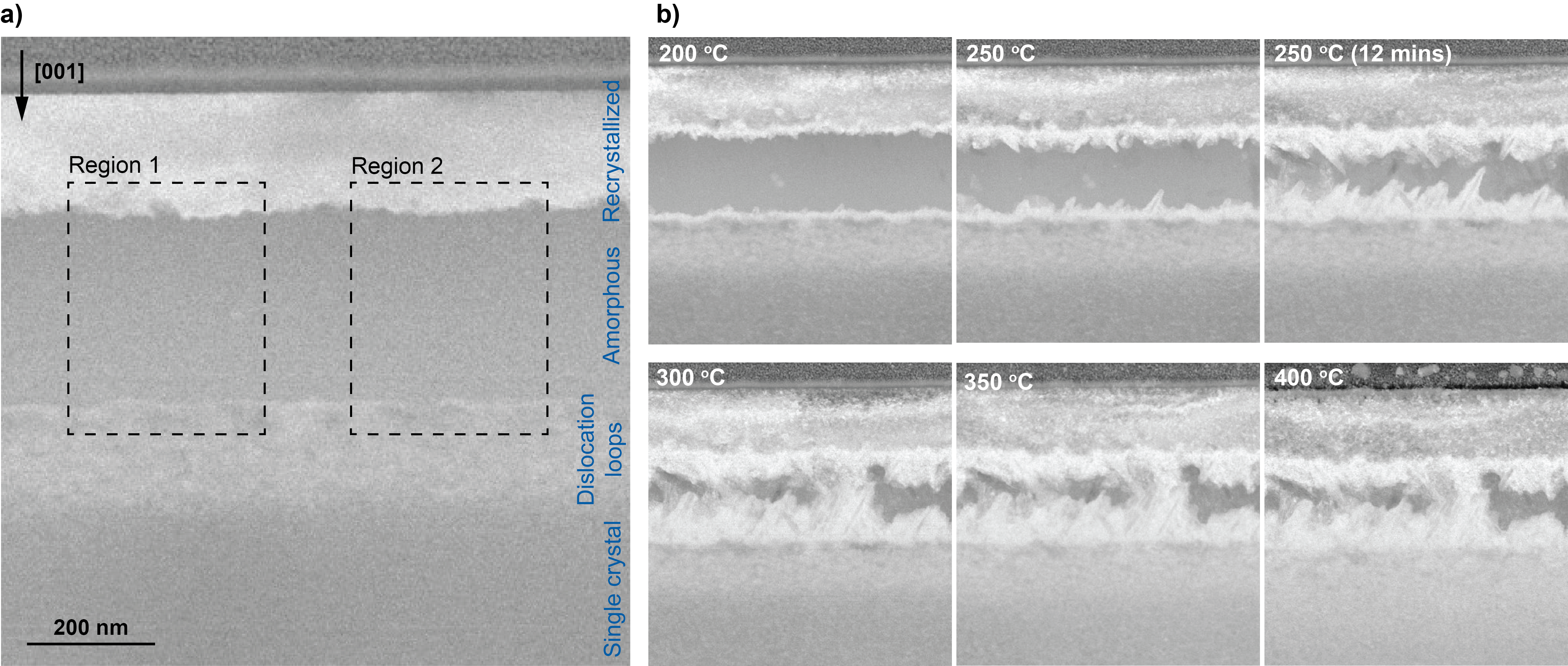}
  \caption{(a) GaAs irradiated with 400~keV Ne$^{++}$ ions exhibits an amorphous layer, with partial recrystallization occurring within the top approximately 200~nm. This recrystallization progresses over time at room temperature and closely matches the crystalline orientation of the pristine single-crystal GaAs. Beneath the recrystallized region lies an amorphous layer approximately 200~nm thick, followed by a zone containing dislocation loops embedded in the underlying pristine GaAs. 4D-STEM data was collected from the two indicated regions. (b) As the temperature increases, the amorphous layer begins to recrystallize. At 200\textdegree{}C, the crystalline--amorphous interface advances into the amorphous layer. By 250\textdegree{}C, nanocrystalline twinning becomes the dominant recrystallization mechanism, occurring along the \{111\} family of planes. Observations over a 12-minute interval at 250\textdegree{}C reveal that recrystallization dynamics are time-dependent in addition to being temperature-dependent. Panels in (b) are from the same field of view as the region in (a). }
  \label{fig:haadf}
\end{figure*}

Our method applies angular cross-correlation analysis to diffraction patterns, quantifying rotational symmetry variations across \textit{k}-space and temperature. By analyzing angular intensity distributions and diffraction coherence, we extract spatially resolved signatures of local ordering, orientation bias, and defect accumulation associated with the amorphous-to-crystalline transition. These measurements extend beyond traditional $n$-fold symmetry detection to capture nuanced changes in structural order.

Unsupervised clustering (k-means++) of individual diffraction patterns segments the sample into structural classes across the thermal cycle. These classes correlate with angular symmetry and order--disorder maps, allowing us to define thresholds for crystallinity and trace the spatial progression of recrystallization. Notably, three-cluster classification reveals the emergence of a crystal phase that is distinct from both amorphous and pristine GaAs, characterized by lattice distortion, twinning, and elevated defect density. The generalizability of this 4D-STEM approach makes it applicable to a wide range of disordered materials beyond semiconductors, where conventional techniques cannot resolve intermediate-range structural features. By connecting angular symmetry patterns to underlying atomic arrangements, this method enables direct mapping of structural memory in systems lacking long-range periodicity.

However, the evolution of the electronic structure does not coincide fully with that of the atomic structure. Correlative electron energy loss spectroscopy (EELS) reveals that, although the recrystallized GaAs adopts the original [101] out-of-plane orientation (perpendicular to the [001] direction of irradiation) of the single-crystal pristine GaAs, its low-loss electronic structure remains altered. The plasmon peak position and shape differ from both undamaged pristine and amorphous GaAs, suggesting persistent strain, antisite defects, or twin-induced modifications to the band structure. Notably, even after \textit{in situ} annealing up to 400\textdegree{}C in 50\textdegree{}C increments, the electronic structure remains measurably distinct from the undamaged and amorphous states. These findings are corroborated by Doppler-broadened spectroscopy (DBS), a depth-resolved variant of positron annihilation spectroscopy (PAS)~\cite{selim2021positron}, which confirms that recrystallized GaAs, though structurally recovered, retains high levels of point defects.

Together, these results show that amorphous GaAs retains sufficient structural memory to support oriented recrystallization, but that full recovery of functional properties is hindered by residual strain and defects. The ability to detect and map this embedded memory using angular symmetry correlations and unsupervised clustering offers a framework for understanding and engineering recrystallization pathways in structurally complex systems.

\begin{figure*}[t]
\centering
\includegraphics[width = 0.96\textwidth]{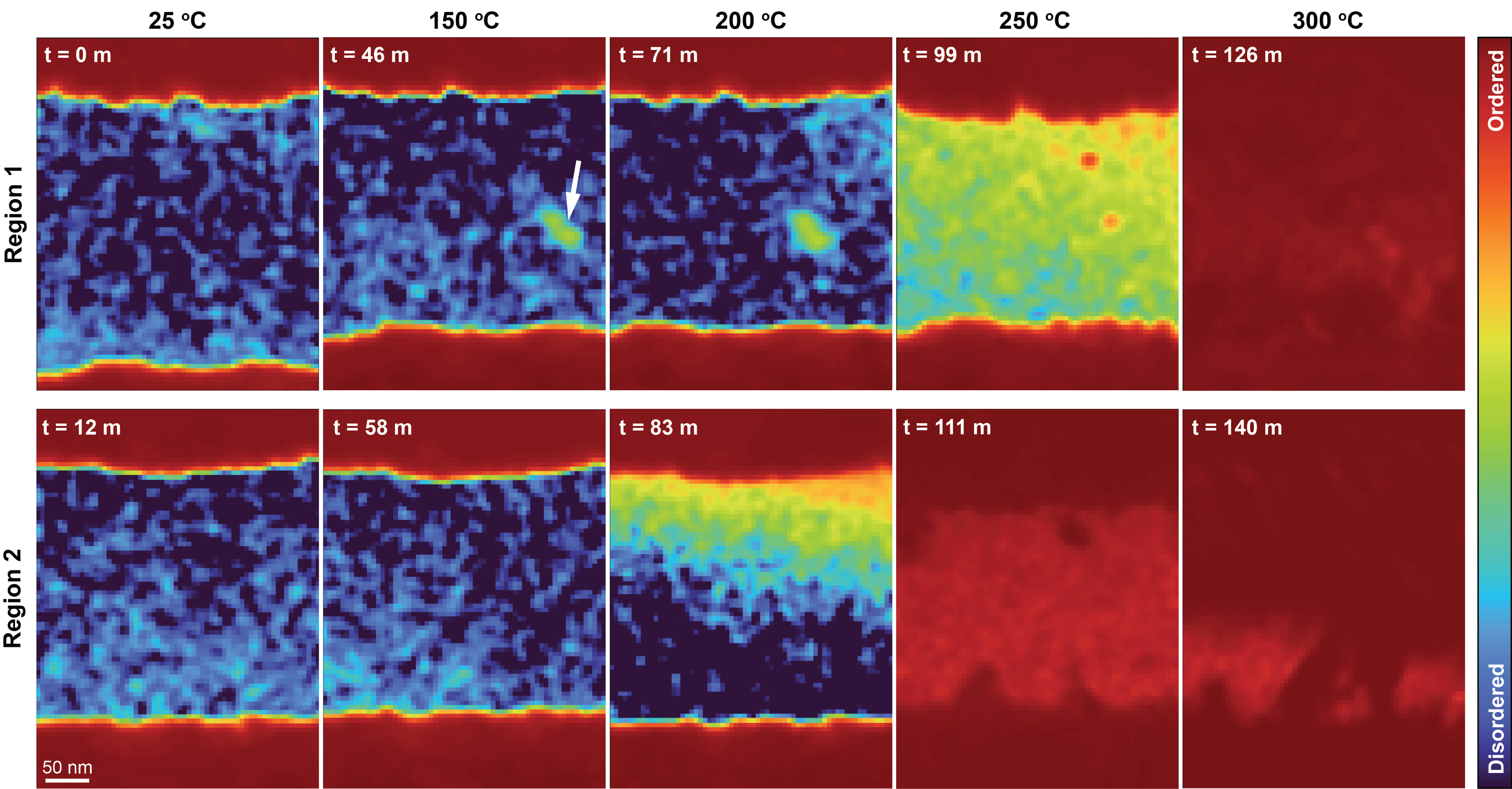}
  \caption{Order--disorder maps from regions 1 and 2, as shown in Fig.~\ref{fig:haadf}a, indicate the relative degree of crystallinity across the sample. Each pixel corresponds to a single probe position and its associated diffraction pattern. The 25\textdegree{}C case shows the distribution of ordering prior to heating. Between 150\textdegree{}C and 200\textdegree{}C, the crystalline fronts of both the bulk and recrystallized regions advance epitaxially into the amorphous layer. At approximately 250\textdegree{}C, twins begin to rapidly extend into the amorphous region, resulting in near-complete recrystallization. These temperature steps were selected because the majority of recrystallization occurs within this range and is complete by 300\textdegree{}C. In Region 1, the more ordered spot appearing at 150\textdegree{}C on the right side of the amorphous layer (indicated with an arrow) was created by the electron beam and served as a fiducial marker. Time steps are shown in the upper left corner of each map; it takes approximately 12 minutes to collect each dataset.}
  \label{fig:ordering}
\end{figure*}

\begin{figure}[t]
\centering
\includegraphics[width = 0.48\textwidth]{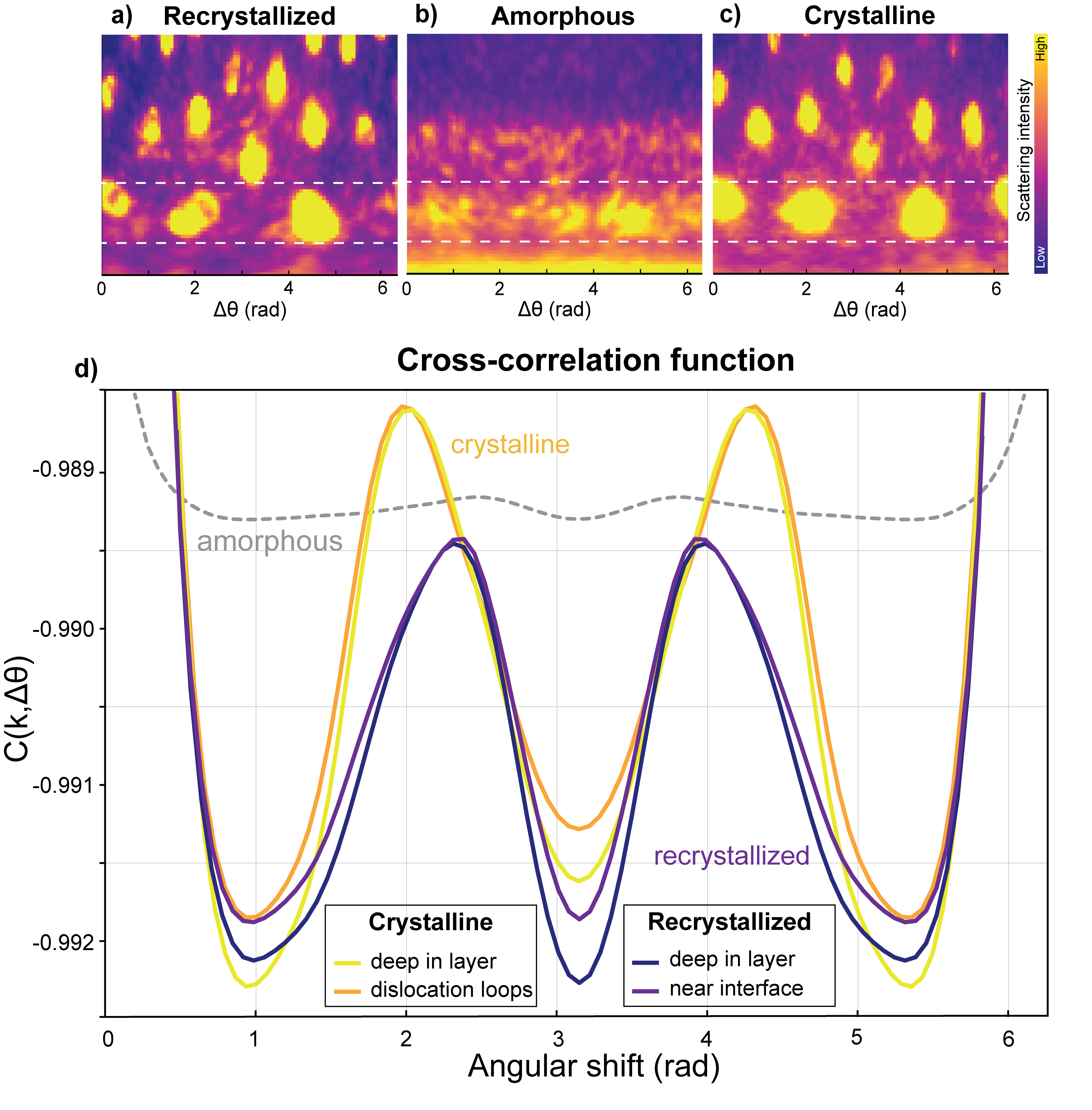}
  \caption{Polar-transformed diffraction patterns from representative regions of (a) the recrystallized layer, (b) the amorphous layer, and (c) the undamaged GaAs at 25\textdegree{}C. Dotted white lines indicate the range of scattering vectors (k) used to compute the cross-correlation function (CCF), encompassing the first diffuse halo in the amorphous region and the 111 diffraction peaks in the crystalline regions. (d) CCF amplitude, C(k,$\Delta\theta$), was calculated as a function of angular shift to quantify the angular periodicity of diffracted intensity at a given spatial frequency. The amorphous region exhibits the lowest CCF amplitude, consistent with a lack of long-range order. Both crystalline regions show dominant two-fold symmetry; however, the recrystallized layer displays broader and less coherent peaks, indicating increased lattice disorder and orientational variation relative to the undamaged crystal. CCF measurements were taken from regions deep within both the recrystallized and undamaged layers, as well as near the amorphous-recrystallized and dislocation loop-amorphous interfaces. In both cases, regions further from the interface exhibit stronger angular periodicity and more defined long-range order.}
  \label{fig:symmetry}
\end{figure}

\section*{Results}

\subsection*{Structural evolution}
Ne$^{++}$ irradiation at 400 keV to a fluence of $2 \times 10^{16}$ ions/cm$^2$ produced a distinct amorphous region in GaAs oriented with the [001] direction normal to the ion source. Immediately following irradiation, the entire top layer was amorphous with no evidence of surface crystallinity (TEM image in Fig. S1e of the Supplemental Information). After a week at ambient conditions, a thin recrystallized layer formed at the surface, leaving an approximately 510 nm amorphous layer (Figure~\ref{fig:overview}a). In addition to this surface regrowth, recrystallization was also observed to initiate from the lower interface above the band of dislocation loops, suggesting that both surfaces of the amorphous region act as nucleation fronts. The lamella surfaces were oriented parallel to the (101) planes when removed from bulk, as shown in Figure S1 of the Supplemental Information. Beneath the amorphous region, a crystalline band containing dense dislocation loops was observed. Strain mapping across the dislocation loops (Figure S2 of the Supplemental Information) reveals that heating relaxes the strain field created by these end-of-range dislocations, suggesting that the dislocations disappear at elevated temperatures. While this evolution is in itself interesting, this work focuses on temperature-induced structural evolution of the amorphous layer. 

Elemental analysis revealed a maximum embedded Ne concentration of approximately 1 at.\%, located at the upper edge of the dislocation-rich crystalline layer (Figure~\ref{fig:overview}a). No visible gas bubbles or voids were detected in either the amorphous or crystalline regions. SRIM simulations of the displacements per atom (dpa) and implanted Ne concentration are shown in Figure~\ref{fig:overview}b.

Prior to \textit{in situ} heating, partial recrystallization was observed at room temperature in samples stored for extended periods post-irradiation. This ambient regrowth produced a layered geometry composed of a roughly 200 nm recrystallized surface layer, followed by a 270 nm amorphous layer, and a $\sim$30 nm recrystallized interface zone adjacent to the crystalline substrate. Recrystallization appeared to initiate from both the free surface and the crystalline interface, with the top-down regrowth being more pronounced in terms of thickness and continuity (Figure~\ref{fig:overview}c).

The partially recrystallized lamella was heated \textit{in situ} from room temperature ($\sim$25 \textdegree{}C) to 400\textdegree{}C in 50\textdegree{}C increments. STEM imaging (Figure~\ref{fig:haadf}) revealed two distinct recrystallization regimes. Between 150\textdegree{}C and 250\textdegree{}C, epitaxial growth progressed slowly into the amorphous layer from both interfaces, maintaining alignment with the undamaged crystal orientation along [101]. At 250\textdegree{}C and above, recrystallization accelerated and was dominated by the formation of dense \{111\} twin networks. These twins formed along the \{111\} planes of the original crystal and preserved the global orientation of the initial structure. By 300\textdegree{}C, the amorphous layer was fully recrystallized across the observed regions.

Order--disorder maps were generated from 4D-STEM datasets collected from regions 1 and 2 (Figure~\ref{fig:ordering}). These maps visualize the spatial distribution of structural ordering and highlight transitions between amorphous and crystalline regions. The blue-green regions correspond to areas lacking long-range order, defined operationally here as amorphous. In Region 1, the spot of relatively higher ordering within the amorphous layer, indicated by the arrow, was created as a fiducial marked by holding the electron beam in place.

At 150\textdegree{}C, early-stage epitaxial recrystallization was observed at the bottom crystalline interfaces, with recrystallized regions extending approximately 5-10 nm into the amorphous layer. By 200\textdegree{}C, recrystallization initiated from the top surface and advanced downward into the amorphous region. At 250\textdegree{}C, significant additional growth occurred, and by 300\textdegree{}C, recrystallization appeared complete across both regions 1 and 2. A full temperature series of order--disorder maps for Region 1 is provided in Figure S3 of the Supplemental Information.

To assess local symmetry and long-range order, angular cross-correlation functions (CCFs) were computed from polar-transformed diffraction patterns. The CCF quantifies angular periodicity in the scattering intensity, providing a measure of orientational order and structural coherence~\cite{wochner2009xray,liu2015interpretation}. Figure~\ref{fig:symmetry} shows representative polar transformations from the recrystallized layer, the amorphous layer, and the pristine substrate, along with corresponding CCF intensity profiles. The amorphous region exhibited the lowest CCF amplitude, consistent with minimal long-range order~\cite{kennedy2025mapping}. Both the undamaged and recrystallized regions showed dominant two-fold symmetry in the $k$-range corresponding to \{111\} reflections, consistent with the [101] zone axis where opposing (111) Bragg reflections produce angular periodicity at 180\textdegree{} (2-fold symmetry). The CCF peaks in the undamaged region were sharp and high in amplitude, indicating well-defined orientational order and low local disorder. In contrast, the recrystallized region showed broader, lower-intensity peaks with visible shoulder features, reflecting increased angular spread and a distribution of local environments. The decreased spacing of CCF peaks from 2.27 rad in undamaged GaAs to 1.54 rad in the recrystallized layer quantifies enhanced angular disorder and suggests lattice strain or mosaicity introduced during regrowth. 

Figure~\ref{fig:symmetrymap} shows $n$-fold angular correlation maps generated on a pattern-by-pattern basis. Comparison between unsupervised clustering (Figure~\ref{fig:clustering}) and the order–disorder and symmetry maps shows that clustering is sensitive to the types of ordering (symmetries), not just the degree of ordering.

\subsection*{Functional evolution}

Low-loss EELS were collected from four areas at 25\textdegree{}C: the recrystallized layer, the amorphous layer, the dislocation loop region, and the undamaged bulk (Figure~\ref{fig:eels}). The undamaged crystal blelow the dislocation loops exhibited a plasmon peak centered at 17.1 eV, a secondary feature at 22.6 eV attributed to Ga 3\textit{d} to conduction band transitions, and a smooth onset corresponding to the band gap tail. The plasmon peak, which reflects the collective oscillation of valence electrons, was suppressed in intensity and shifted to 17.6 eV in the recrystallized layer. The amorphous and dislocation loop regions also showed reduced plasmon intensities, with peaks at 17.0 eV and 17.1 eV, respectively. Given the 0.2 eV energy resolution, the plasmon peaks in the amorphous and dislocation loop regions are considered equivalent. Both the undamaged and dislocation loop regions exhibited small edge-like features near 5.5 eV, while the amorphous and recrystallized layers showed broader, shelf-like responses in this range, indicating a range of transition pathways.

DBS measurements, shown in Fig.~\ref{fig:dbs}, were taken using a variable-energy positron system for four GaAs [001] samples. $S$ (low-momentum electron-positron annihilations) and $W$ (high-momentum annihilations) parameters were measured as a function of implantation energy for a pristine undamaged wafer, the irradiated wafer, and two different annealed samples (one annealed to 150\textdegree{}C for 45 minutes, and the other at 300\textdegree{}C for five minutes). Implanted positrons are sensitive to the presence of vacancy-like defects, which act as scattering and trapping centers for the thermalized positrons. Trapping into defects results in a decrease in the overlap of the positron wavefunction with the tightly bound core electrons, leading to a narrowing of the 511 keV annihilation line increasing $S$ and decreasing $W$. Through VEPFIT analysis~\cite{vanveen1995vepfit} information about the positron diffusion length and effective thickness of the distinct
layers of a target system can be extracted, providing some estimate of the defect
concentration~\cite{corbel1988positron}. The undamaged GaAs sample shows a smooth transition from surface to bulk annihilation with bulk-like $S$ and $W$ values, while the irradiated sample exhibits a narrow defect-rich layer near the surface with a very short positron diffusion length indicative of a high vacancy concentration and dislocation-rich regions with near-bulk $S$ and $W$ values. The two annealed samples display expanded defect-rich regions compared with the irradiated sample, consistent with imperfect recrystallization.

\section*{Discussion}

\subsection*{Structural memory embedded in amorphous heterogeneity}

The degree of structural disorder in the amorphous GaAs layer varies slightly with depth, following the displacement damage profile generated during irradiation, as indicated by the gradient in the blue amorphous region in (Figure~\ref{fig:ordering}). The region corresponding to the peak dpa exhibits the most extensive amorphization, consistent with expectations that high displacement damage more severely disrupts long-range atomic order. In contrast, near the surface, where the dpa is lower, a greater degree of short- and medium-range order is retained, giving rise to paracrystalline features embedded within the amorphous matrix. Here, we define short-range ordering as involving the first coordination shell, and medium-range ordering as encompassing approximately the first 3–4 coordination shells, corresponding to a length scale of roughly 0.5 to 5 nm. This preserved structural framework likely serves as an ersatz template for recrystallization, facilitating the top-down recovery observed in the irradiated GaAs. This is in contrast to the recrystallization at the lower interface, which has a crystalline template from which to grow. Additionally, enhanced defect mobility and surface diffusion near the free surface promote atomic rearrangement and likely contribute to the initiation of recrystallization at the surface.

\begin{figure}[t]
\centering
\includegraphics[width = 0.48\textwidth]{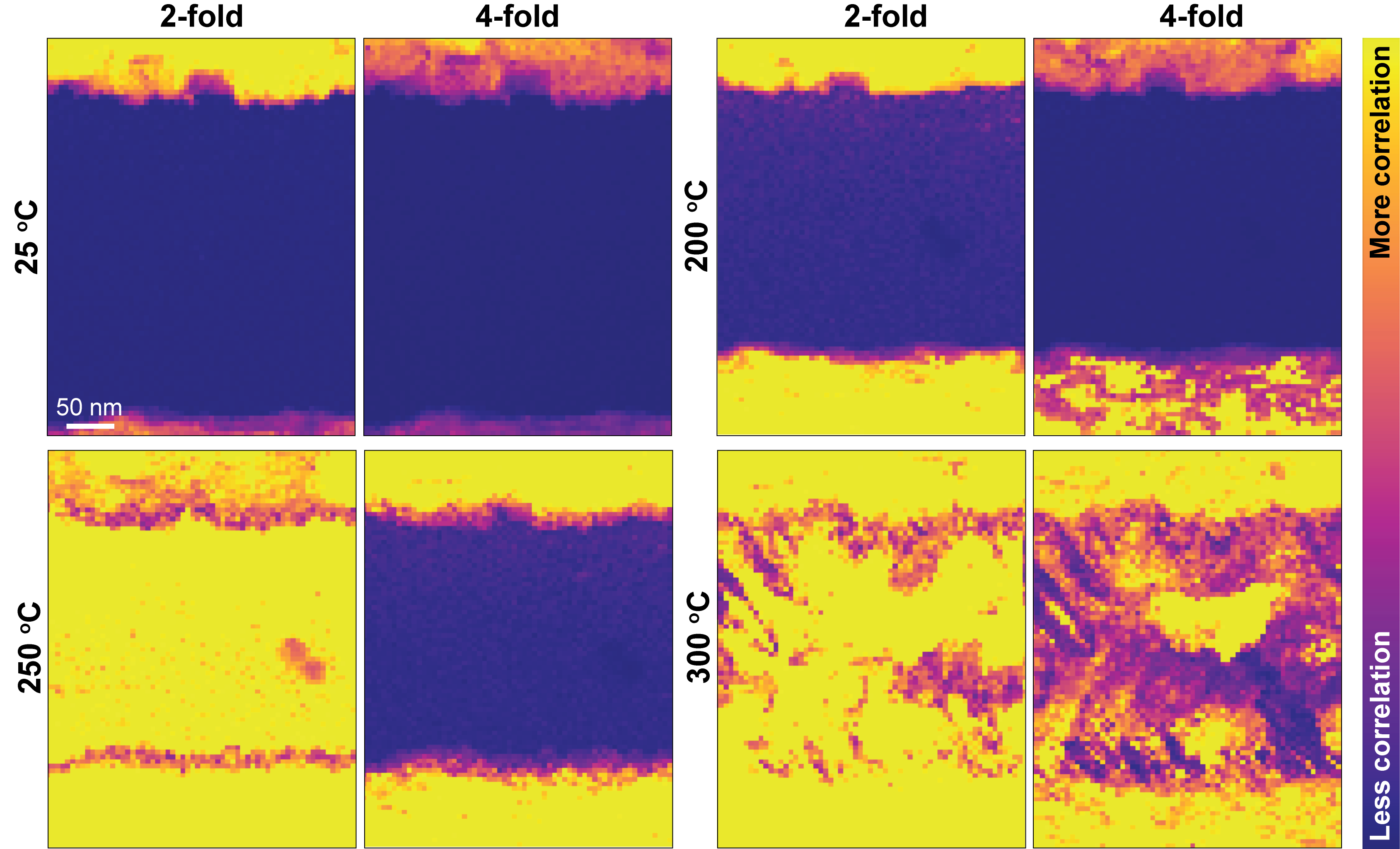}
  \caption{{N}-fold angular correlation maps are generated from 4D-STEM datasets of region 1 by measuring periodicity in each diffraction pattern. The scattering vector range extends beyond the \{111\} reflections, capturing higher-order angular correlations. Shown are 2- and 4-fold maps across the thermal regime in which \{111\} nanotwins nucleate and grow from crystalline regions into the amorphous layer, leading to near-complete recrystallization. Room temperature maps are shown for comparison. At 200\textdegree{}C, rotational symmetry signals increase in crystalline regions. At 250\textdegree{}C, 2-fold symmetry becomes more prevalent across the field of view, including the amorphous layer, though it remains suppressed at the amorphous--crystalline interfaces. By 300\textdegree{}C, twinning extends through most of the amorphous region. The abrupt increase in 2-fold symmetry suggests a structural transition driven by twin network formation. These twins produce streaked diffraction patterns with dominant 2-fold symmetry, explaining the lack of a corresponding rise in 4-fold symmetry. A single polycrystalline pocket (bright yellow in 4-fold maps) persists in the now-recrystallized region; amorphous content is negligible. Differences in angular correlation strength between the recrystallized and undamaged regions arise from slight misorientation, increased lattice distortion, and higher defect densities. Angular correlation maps for region 2 are shown in Figure S4 of the Supplemental Information.}
  \label{fig:symmetrymap}
\end{figure}

The order--disorder maps in Figure~\ref{fig:ordering} reveal depth-dependent variations in disorder within the amorphous layer, with regions of slightly increased ordering (light blue) appearing near crystalline interfaces. These regions remain amorphous but exhibit more heterogeneous Bragg scattering, indicating some residual structural coherence. In other words, these regions exhibit greater degrees of paracrystallinity than the more fully amoprhous regions, suggesting that the recrystallization occurs not by an abrupt amorphous-crystalline transition, but by a more gradual process that includes intermediate levels of structural disorder, akin to our previous work showing intermediate levels of chemical disorder at the recrystallization front~\cite{kennedy2024insights}.

The recrystallized layer adopts the same crystal structure and orientation as the pristine GaAs, with minimal misorientation (less than 1\textdegree{}) measured from shifts in 4D-STEM diffraction patterns. This structural coherence is consistent with residual paracrystallinity acting as a recrystallization template, suggesting that the paracrystalline regions preserve their initial orientation rather than rotating during disordering. The small misorientation also accounts for the contrast difference observed in HAADF-STEM imaging (Figure~\ref{fig:haadf}), where the recrystallized layer appears brighter due to slight changes in diffraction conditions that direct more scattered electrons into the high-angle detector range.

From 25\textdegree{}C to 200\textdegree{}C, a layer of GaAs with a retained amorphous structure remains beneath the recrystallized layer, aligned with the depth of maximum dpa. This portion exhibits the greatest disruption of periodicity and lacks the paracrystalline framework necessary to nucleate crystalline order, preventing recrystallization under the same conditions. These observations are consistent with prior studies~\cite{shao2014effect,li2023quantitative,kennedy2024insights} showing that ion irradiation can induce depth-dependent disorder, where surface effects, such as enhanced defect recombination, stress relaxation, and injected interstitial gradients, significantly impact local defect evolution and subsequent microstructural recovery. The presence of paracrystalline order in GaAs may reflect a broader principle in disordered network solids: that nominally amorphous structures can host localized, energetically favorable crystalline-like domains~\cite{rosset2025signatures,treacy2012local}. This intermediate regime is neither fully disordered nor long-range ordered and may serve as a precursor state guiding phase transitions, as recently proposed for amorphous silicon~\cite{rosset2025signatures,nakayama2025symmetry} and documented in amorphous oxides~\cite{niu2023structural}. This is best understood as a two-phase microstructure that cannot be characterized by a single motif.

The top-down recrystallization observed here occurred at room temperature over the course of approximately one month. This low-temperature recovery was likely enabled by the thin geometry of the TEM lamella, which provides additional high-mobility surfaces to promote atomic rearrangement. EELS \textit{t/$\lambda$} measurements estimate the thickness of the region from which 4D-STEM was collected as between 62 nm (near the surface) and 78 nm (in the crystalline dislocation loop layer). Thus, the recrystallization process appears to result from a combination of structural memory encoded in nanometer-scale ordered regions, sufficient time at ambient conditions, and the enhanced kinetics afforded by reduced dimensionality.

\subsection*{Two recrystallization regimes}
As discussed, two recrystallization regimes are observed in GaAs. The first is a lower-temperature regime characterized by the slow propagation of the crystalline front into the amorphous layer. The second is a more dynamic regime, marked by the formation and partial coalescence of a nanotwin network along \{111\} planes. In both regimes, the orientation of the undamaged GaAs, which corresponds to the orientation of the amorphous region prior to irradiation, is preserved.

The transition between these regimes occurs at approximately 250\textdegree{}C, although some evidence of rapid recrystallization is also observed when the sample is held at 200\textdegree{}C for extended periods, as shown in Figure~\ref{fig:ordering}, Region 2, in the 200\textdegree{}C panel. As noted above, the recrystallization temperature for bulk amorphous or damaged GaAs is typically reported around 300-400\textdegree{}C, but values as low as 177\textdegree{}C have been cited in thin-film geometries~\cite{schulz1996optimum,desnica-frankovic1999different,herold1989investigation}. Thus, the low-temperature recrystallization observed here falls below the conventional range and is likely attributable to the thin geometry of the FIB lamella, remnant paracrystallinity, and extended annealing time. Similarly, in germanium, ionization-induced recovery has been demonstrated as a nonthermal route to restore crystallinity at room temperature. This process reveals that crystalline-to-amorphous transitions are reversible, challenging the long-standing view that thermal energy is required for recovery~\cite{velisa2025revealing}.

\begin{figure}[t]
\centering
\includegraphics[width = 0.48\textwidth]{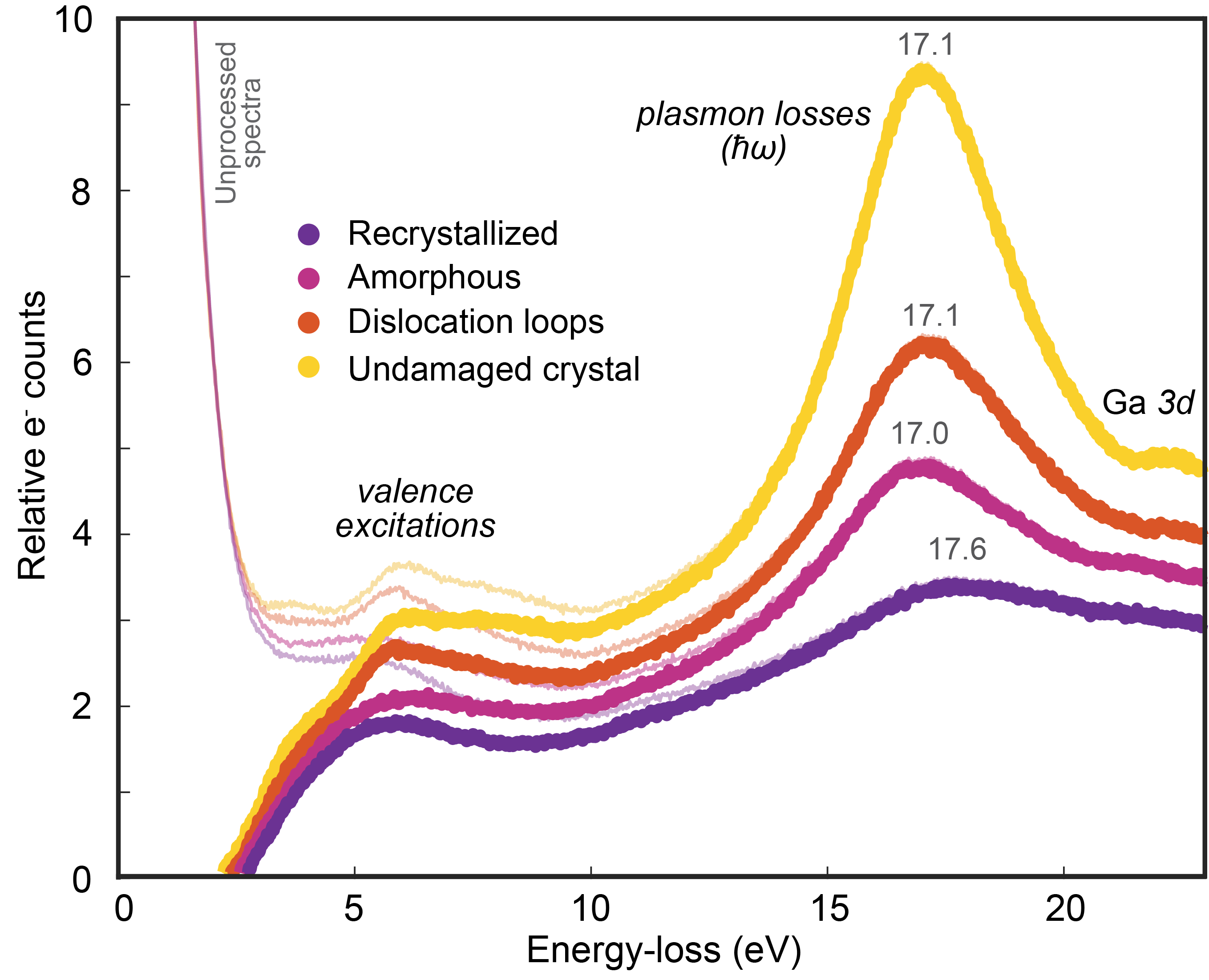}
  \caption{EELS spectra collected from various depths beneath the GaAs surface, including the recrystallized layer, the amorphous layer, the dislocation loop region, and the undamaged bulk. Raw spectra are shown as transparent lines; background-subtracted and plural-scattering-corrected spectra are shown as opaque lines. In the recrystallized layer, the plasmon peak is substantially suppressed in intensity and shifted to 17.6 eV, consistent with lattice disorder and plasmon dampening. The amorphous and dislocation loop regions also exhibit reduced plasmon peak intensities compared to the undamaged crystal, with their peaks shifted downward to 17.0 and 17.1 eV, respectively, suggesting a lower effective density or increased structural disorder. Both the undamaged and dislocation loop regions show small edge-like features near 5.5 eV, potentially corresponding to interband or valence excitations. In contrast, the recrystallized and amorphous regions exhibit a broader, shelf-like feature in this range rather than a distinct peak.
 }
  \label{fig:eels}
\end{figure}

Over roughly four weeks at ambient temperature, the top of the amorphous region recrystallizes in nearly the same orientation as the original GaAs. According to SRIM simulations (Figure~\ref{fig:overview}b), the dpa near the surface of the damaged region is significantly lower than in the center and lower-end-of-range regions of the amorphous layer. Together, these observations suggest that the reduced dpa near the surface permits the retention of short- and medium-range order that cannot be detected using conventional diffraction techniques. At low temperatures and over long timescales, these paracrystalline motifs serve as templates for recrystallization, slowly expanding and promoting long-range structural reordering.

This phenomenon of paracrystalline-guided ordering appears active at low temperatures, where thermal diffusion is limited and structural memory plays a more dominant role in guiding atomic rearrangement~\cite{velisa2025revealing}. In relatively simple two-component systems, the energetic favorability of an ordered lattice strongly promotes recrystallization once sufficient atomic mobility is available, in contrast to multi-component systems such as bulk metallic glasses, where chemical and topological complexity assist in stabilizing the disordered state.

The colormap scaling in Figure~\ref{fig:ordering} was specifically chosen to emphasize variations in the degree of ordering within the amorphous region. At room temperature and up to 250\textdegree{}C, subtle heterogeneity is observed, with slightly higher ordering, lighter blue shades, evident near the crystalline–amorphous interfaces. This suggests that short- or medium-range structural remnants form closer to these boundaries. Additionally, the order and disorder maps (Fig.~\ref{fig:ordering})  reveal that the recrystallized layer near the surface is slightly more faceted and rougher than the lower, more uniform crystalline region. In both Regions 1 and 2, recrystallization at elevated temperatures is observed to initiate from the top right corner and propagate downward. The spatial asymmetry favoring the right side is likely a result of a slight time delay during 4D-STEM scanning, which proceeds in a raster pattern from left to right and top to bottom. This subtle time offset may enable localized thermal effects to accumulate in the rightmost regions first.

The observation that re-ordering initiates at both the top and bottom of the amorphous layer, but occurs preferentially at the interface between the amorphous layer and the already recrystallized surface, indicates that properties of the interface or nearby amorphous region strongly influence recrystallization behavior above approximately 250\textdegree{}C. We propose two contributing mechanisms for this. First, the increased surface roughness of the recrystallized layer likely provides favorable heterogeneous nucleation sites, particularly for twinning. This phenomenon is well documented across material systems, where roughness and surface features significantly lower the nucleation barrier for specific crystallographic orientations or defect structures~\cite{janish2020insitu,zhang2025study}. Second, structural variations and embedded motifs within the amorphous matrix, such as remnants of Ga–As tetrahedral coordination or short-range ordered clusters~\cite{liu2025review}, may influence nucleation and growth kinetics by providing partial crystalline templates or lowering the local free energy barrier for ordering.

Supporting this hypothesis, the symmetry maps shown in Figure~\ref{fig:symmetrymap} and Figure S4 of the Supplemental Information indicate increased two-fold rotational symmetry near the top of the amorphous region, especially around 200\textdegree{}C, compared to the bottom of the amorphous layer. This increase in local symmetry is not captured in the order–disorder maps, which quantify only relative crystalline signal. The presence of specific rotational symmetries near the surface suggests that partially ordered paracrystalline configurations, undetectable by traditional diffraction but revealed in the symmetry analysis, may play a critical role in guiding recrystallization. The medium- and short-range ordering serve as guides for recrystallization, and the high atomic mobility associated with amorphous structures ~\cite{ridley2024amorphous,kim2012uranium,kreller2019massively} enables rapid recrystallization across the entire region, as evidenced in the HAADF-STEM images in Figure~\ref{fig:haadf}b.

These observations point toward a recrystallization process not solely governed by temperature, but also heavily influenced by local structural remnants, interface morphology, and symmetry motifs within the amorphous layer. These factors jointly determine the directionality, mechanism, and outcome of recrystallization under varying thermal conditions.

\subsection*{Symmetry mapping on the nanometer-length scale}
Pattern-by-pattern symmetry mapping provides nanometer-scale insight into angular correlations within both the crystalline and amorphous regions of GaAs. The cross-correlation functions (CCFs) in Figure~\ref{fig:symmetry} are calculated over the reciprocal space vector (\textit{k}-vector) range corresponding to the [111] Bragg reflections and the amorphous halo. Comparing Figures~\ref{fig:symmetry}a and c, we observe that even between the two crystalline regions, recrystallized and initially undamaged, there is greater variation and increased signal beyond the Bragg peaks in the recrystallized pattern. This is quantified in the line plot in Figure~\ref{fig:symmetry}, where the CCF peaks in the recrystallized region show lower amplitude and broader angular distributions. The shoulder features in the recrystallized spectra align with the primary peaks of the initial GaAs structure, indicating a preserved lattice orientation but with increased structural disorder and a higher density of defects. While long-range order is largely recovered after recrystallization, strain fields and point defects remain.

\begin{figure}[t]
\centering
\includegraphics[width = 0.48\textwidth]{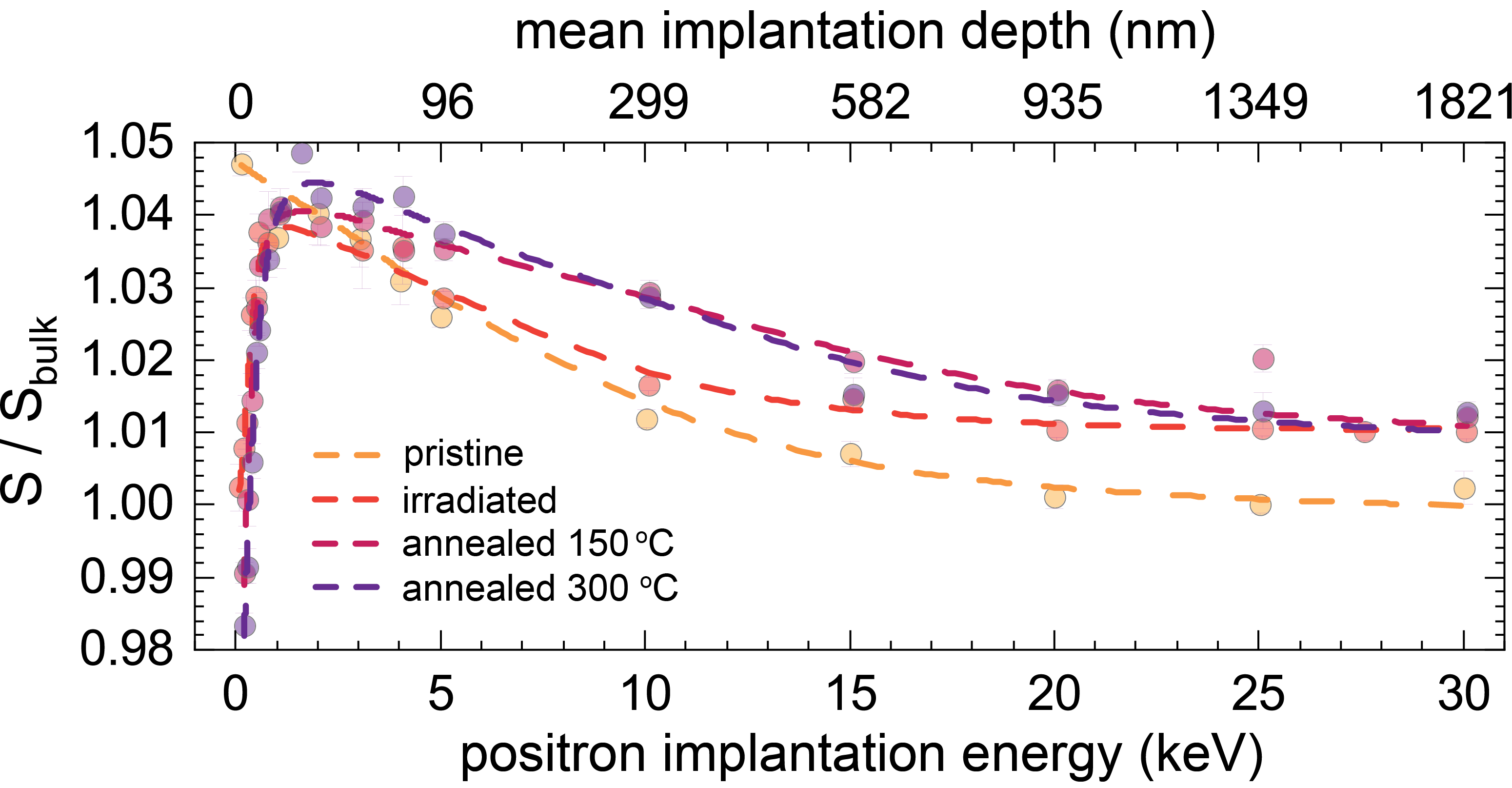}
  \caption{Positron DBS measurements of GaAs [001] targets are plotted for a pristine undamaged  wafer, an irradiated wafer, an irradiated wafer annealed at 150\textdegree{}C for 45 minutes, and an irradiated wafer annealed at 300\textdegree{}C for five minutes. Dashed curves illustrate the result of VEPFIT analyses from the the DBS data (filled circles). Top x-axes indicate the calculated mean positron implantation depth as a function of energy~\cite{jones2025positron}, which varies between the targets due to the varying densities following irradiation and annealing.
 }
  \label{fig:dbs}
\end{figure}

In both crystalline regions, 2-fold symmetry dominates the rotational correlations, as expected for electron beam incidence along the [101] zone axis of the GaAs lattice. In contrast, the amorphous region exhibits significantly reduced angular correlation and lacks consistent 2-fold symmetry, as detailed in Figure S5 of the Supplemental Information. Instead, the amorphous structure displays broader, weaker angular correlations, with contributions from multiple 2-fold and 4-fold components, consistent with short- and medium-range structural motifs that lack long-range periodicity.

The symmetry maps in Figure~\ref{fig:symmetrymap} were generated using an extended range in reciprocal space, capturing angular periodicity beyond the [111] peaks alone. Both 2-fold and 4-fold maps are shown to illustrate the progression from amorphous to crystalline structure with increasing temperature. At 200\textdegree{}C, a modest increase in 2-fold rotational correlation appears near the top of the amorphous region, adjacent to the recrystallized surface. This supports the interpretation in Figure~\ref{fig:ordering} that recrystallization initiates in a top-down fashion. In this context, the 4-fold symmetry channel becomes particularly useful for identifying regions of polycrystallinity or twinning that deviate from the ideal \{101\} orientation. For instance, at 200\textdegree{}C, a distinct region of increased 4-fold symmetry appears in the center of the amorphous layer, corresponding to a GaAs pocket with approximate \{101\} orientation but with structural disorder.

These angular correlation analyses probe a broader portion of reciprocal space than conventional Bragg peak tracking, providing a more inclusive metric of local periodicity. As a result, 2-fold symmetry offers a relatively stringent criterion for identifying truly ordered regions, while 4-fold symmetry captures broader measures of crystallinity and structural coherence.

\begin{figure*}[!ht]
\centering
\includegraphics[width=0.98\textwidth]{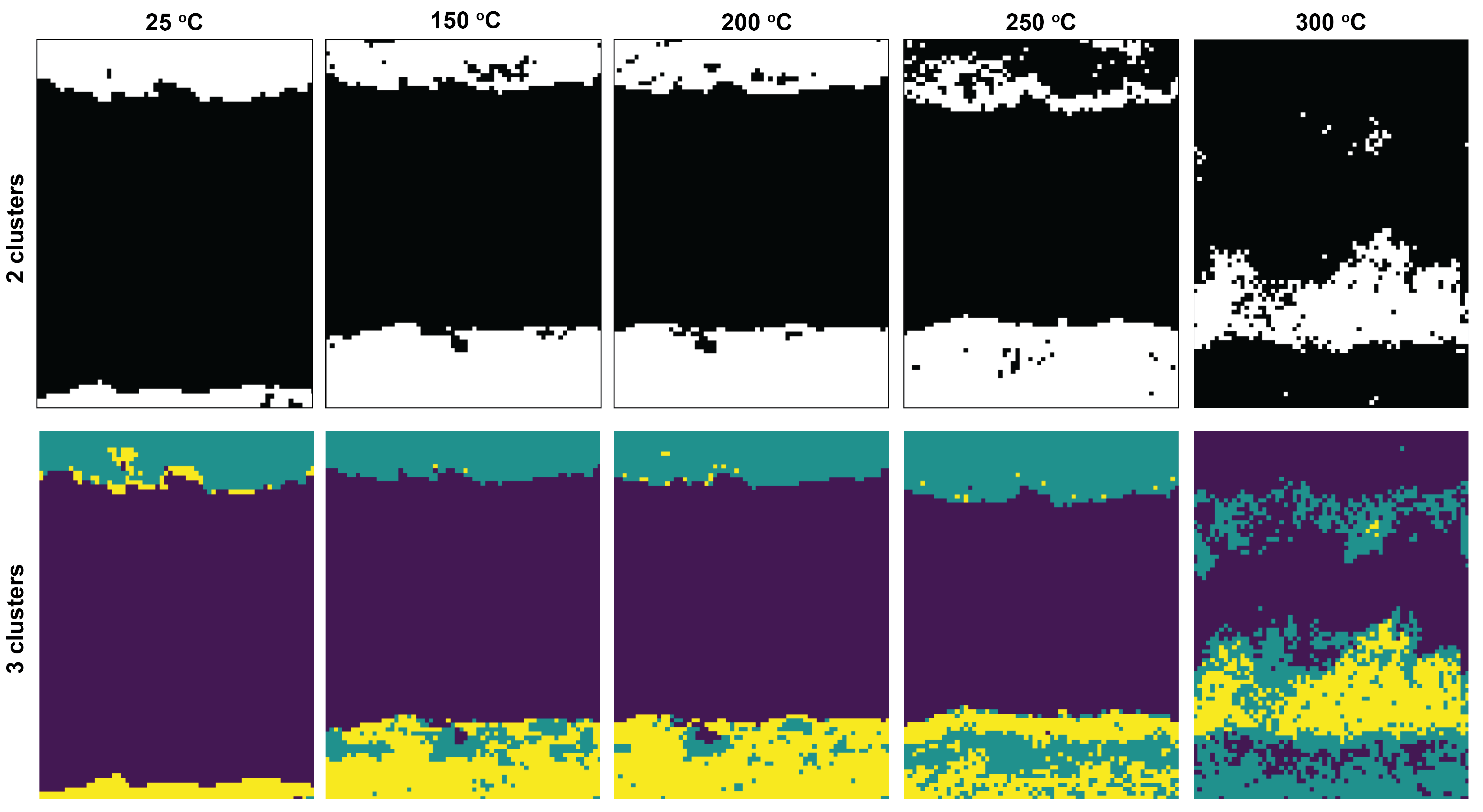}
\caption{\label{fig:clustering} Diffraction patterns from Region 1 were classified using the k-means++ algorithm into two and three clusters. In the two-cluster classification, amorphous and crystalline regions are separated, indicating that features associated with the degree of crystallinity (such as net pattern intensity and the presence of peaks versus an amorphous halo) were used for clustering. This trend holds up to 250\textdegree{}C, above which the clustering begins to distinguish between crystalline patterns with and without twinning cross-hatching. Three-cluster classification was also performed to allow for the identification of an additional structural class. For temperatures at and below 200\textdegree{}C, the clusters appear to correspond predominantly (though not exclusively) to amorphous, pristine crystalline, and recrystallized regions. Within the crystalline regions, some mixing occurs, with pixels classified as both recrystallized and initial crystal. At 250\textdegree{}C, the classification still separates amorphous from crystalline patterns, but within the lower crystalline region, two distinct classes emerge. One of these, corresponding to a recrystallized front (teal color), is distinct from the top recrystallized layer. At 300\textdegree{}C, the three-cluster classification no longer includes an amorphous class. Instead, it identifies: (1) a single-crystal region with weaker signal (purple), (2) a class resembling the initial undamaged structure (teal), and (3) a class corresponding to strong twinning (yellow).}
\end{figure*}

Interestingly, at 250\textdegree{}C, the interface between the previously amorphous layer and the recrystallized region exhibits reduced 2-fold angular correlation compared to either region alone. This suggests that although the amorphous region has recrystallized, a degree of interfacial disorder persists, indicating that the recrystallization process is not strictly epitaxial, even though it initiates from a crystalline template. Additional symmetry maps in Figure S4 of the Supplemental Information from Region 2 further support this interpretation, showing the onset of twinning and increased rotational symmetry near 250\textdegree{}C.

Unsupervised clustering was performed to evaluate the distinction between diffraction patterns corresponding to different structures. Comparing the symmetry maps to the k-means++ cluster maps (Fig.~\ref{fig:clustering}) reveals strong alignment between angular correlation patterns and clustering outcomes. In this approach, k-means++ groups diffraction patterns into clusters (bins) by minimizing differences in their angular cross-correlation signatures, with improved initialization to avoid biased seeding. At lower temperatures, the 2-cluster classification primarily separates amorphous and crystalline regions, while at higher temperatures it differentiates between single-crystal and twinned or polycrystalline domains. The 3-cluster classification provides greater nuance by dividing the data into three bins with distinct angular correlation characteristics. At lower temperatures, this separates undamaged crystal, recrystallized regions, and amorphous material, while at elevated temperatures it resolves twinned, distorted, and relatively defect-free crystalline domains. This method provides an autonomous method to map the arrangement of coherent structures at small length scales that can be directly related to measurable properties such as disorder or phase~\cite{kim2025operando,yoo2024unsupervised}. Representative mean diffraction patterns for each bin, which illustrate the basis for the classification, are provided for the ``3 clusters" maps in Figure S6 of the Supplemental Information.

This clustering analysis also identifies subtle heterogeneities within the recrystallized regions. Specifically, it distinguishes areas of increased disorder, lattice distortion, and strain that are less evident in the order--disorder maps. This suggests that symmetry-based correlation and unsupervised clustering together provide a more comprehensive view of the recrystallization process than intensity-based metrics alone. While order--disorder maps quantify the amount of crystalline signal, angular correlation and clustering reveal how that order is spatially structured and disrupted.

\subsection*{Spectroscopic characterization of defect-driven electronic changes}
Evaluating how structural damage and recrystallization affect the electronic properties of GaAs, low-loss EELS and DBS were employed to probe defect concentrations and the resulting electronic structure modifications. Together, these techniques confirm that although the recrystallized GaAs layer appears structurally coherent in the order–disorder analyses, it remains electronically compromised. This contrast highlights the extreme sensitivity of GaAs functionality to point defects and lattice disorder, even when the long-range crystalline framework has largely reformed.

As shown in Figure~\ref{fig:eels}, the low-loss EELS spectra from the undamaged bulk region exhibit the characteristic response of crystalline GaAs: a strong, sharp plasmon peak centered at 17.1 eV, a smaller secondary peak at 22.6 eV corresponding to Ga 3\textit{d} core-level transitions, and a steep band-edge onset associated with the optical gap. In contrast, the recrystallized region shows two notable deviations. First, the plasmon peak is both damped and broadened, with its maximum shifted to 17.6 eV. This suggests a combination of increased electron scattering, decreased coherence of free-electron oscillations, and potential changes in electron density~\cite{batson2004electron,foerster2019plasmon}. These effects are consistent with a high concentration of point defects, which can trap carriers, scatter plasmons, and alter the dielectric response~\cite{foerster2019plasmon,aguillon2021plasmons,kennedy2024insights}. The energy upshift in the plasmon peak may additionally reflect reduced electronic screening or local strain fields arising from residual lattice disorder.

Second, the band-edge region in the recrystallized GaAs displays a broad, shelf-like shape without the prominent valence excitation feature observed around 5.5 eV in the undamaged and dislocation loops regions. In the undamaged and dislocation-rich crystalline regions, the feature at 5.5 eV is indicative of interband or valence-to-conduction transitions~\cite{batson2004electron}. This feature is negligible in both the recrystallized and amorphous regions, suggesting the presence of localized defect states that smear the density of states near the band edge. The broadened edge and suppressed feature at 5.5 eV are indicative of increased electronic disorder, consistent with tail states introduced by lattice defects~\cite{shuk1997,singh2023impact}.

While amorphous GaAs lacks long-range periodicity, its plasmon peak remains centered at 17.1 eV, similar to that of the undamaged crystal, although with reduced amplitude. This suggests that amorphization disrupts structural coherence without significantly changing the average electron density. In contrast, the recrystallized region, despite being structurally more ordered, exhibits greater plasmon damping and a higher peak energy. This somewhat surprising finding indicates that point defects introduced during recrystallization have a stronger impact on collective electronic behavior than the loss of long-range order.

The dislocation loop region also shows a damped plasmon response but with minimal energy shift, implying an intermediate level of disorder. Overall, the plasmon behavior follows a clear trend: strong and sharp in the undamaged crystal, weakened in the dislocated regions, further damped in the amorphous material, and most severely affected in the recrystallized zone. This pattern highlights the critical role of point defect concentration and distribution in determining electronic response. 

Although the undamaged GaAs band gap is approximately 1.42 eV, direct measurement by EELS is limited by the instrument’s energy resolution and the presence of a long zero-loss tail~\cite{iles2004photovoltaic}. While the shape of the band edge changes markedly with increasing disorder, it is not possible to definitively resolve any shift in the optical gap itself. However, the extended band tailing observed in both recrystallized and amorphous regions is consistent with an increased density of localized states and disrupted electronic transitions. Our results suggest that while short-range order in amorphous GaAs can sustain collective excitations, the defect-rich recrystallized phase undermines coherence more severely by introducing localized scattering centers. This highlights a broader principle emerging across disordered systems: local order may be sufficient to support coherent electronic responses, but the pathway of recovery can have a greater impact on functionality than the mere loss of long-range periodicity. Comparable behavior has been noted in a range of materials, including conventional semiconductors~\cite{divincenzo1984long,brodsky1985structural} and in more exotic systems, such as materials with topological electronic states~\cite{ciocys2024establishing}.

The spectroscopic results emphasize that structural recovery observed through 4D-STEM and HAADF-STEM does not guarantee full electronic recovery. Defects preserved or generated during recrystallization degrade carrier coherence and increase the density of gap states, ultimately suppressing material performance even in nominally crystalline GaAs. Therefore, structural analyses must be complemented by spectroscopic data to fully evaluate the functional restoration of amorphized semiconductors. In some cases, amorphization itself has a smaller impact on electronic functionality than the subsequent recrystallization, highlighting that strategies for damage control in semiconductors must consider both structural and electronic recovery. Simply annealing amorphous regions is insufficient; preventing amorphization in the first place is critical to preserving device performance.

DBS measurements shown in Fig.~\ref{fig:dbs} support EELS findings on electronic structure. In the undamaged sample the spectrum, plotted in Fig.~\ref{fig:dbs}a, is described by a transition
from thermal and epithermal annihilation at the surface at low implantation energy to
nearly complete bulk annihilation at high energy. The fitted bulk $S$ and $W$ values are
$0.5551 \pm 0.0007$ and $0.0926 \pm 0.0002$ respectively and are used to produce
normalized curves. The fitted diffusion length is in good agreement with the value
reported elsewhere, $196 \pm 12$ nm~\cite{soininen1992high}.

Based on the TEM images of the irradiated sample, the system appears to be comprised
of three layers: (i) an amorphous top layer $\sim 510$ nm thick, with a density of
$\sim 76\%$ that of the pristine undamaged crystal (determined via 4D-STEM analysis), followed by
(ii) a $\sim 200$ nm thick crystalline layer with a high density of dislocation loops, and
(iii) the undamaged bulk crystal. The local maxima and minima in $S$ and $W$ at
$\sim 1$--3~keV indicate a narrow layer near the surface with high vacancy-type defect
content. The short diffusion length in this region, evident in the rapid transition at low implantation energy from the surface state to the high-defect peak, corresponds to a vacancy concentration
between about $10^{18}$--$10^{20}$ cm$^{-3}$. The deeper part of the amorphous layer and
the dislocation-rich band yield $S$ and $W$ values that are nearly the same as those in
the bulk crystal. As dislocation loops do not substantially change the lattice density and,
in the absence of associated vacancy defects, only act as weak positron traps, this is
not surprising. The near parity of the observed $S$ and $W$ values in the lower-density
amorphous region is harder to explain, though this region remains difficult to confidently
characterize due to the limited data set.

Annealing temperatures of 150\textdegree{}C and 300\textdegree{}C were chosen to represent a gradual, mild recovery and a more accelerated process, respectively.The two annealed samples are nearly identical, differing only below 10~keV. In both samples the defect-rich region appears to expand deeper, consistent with the findings of the 4D-STEM analysis which indicated imperfect recrystallization. The $S$ and $W$
parameters observed in the defect-rich regions are indicative that vacancies dominate in these regions. In undoped GaAs both Ga and As monovacancies can act as positron traps when
neutral or negatively charged~\cite{corbel1988positron}. Different vacancy types have distinct lifetimes
and normalized $S/W$ values~\cite{gebauer2000defect}. In the annealed sample at a positron implantation energy of 1--1.5~keV we find $S/S_{\mathrm{bulk}} = 1.0566 \pm 0.0018$ and $W/W_{\mathrm{bulk}} = 0.6465 \pm 0.0083$.

Based on these values we can rule out both As and Ga monovacancies and their
associated complexes, where $W/W_{\mathrm{bulk}}$ lies between 0.92--0.71. The value of
$W/W_{\mathrm{bulk}}$ is in close agreement with that reported for the divacancy
$V_{\mathrm{As}}$--$V_{\mathrm{Ga}}$ (0.62). As the diffusion length indicates a high
concentration of defects, positron implantation in the defect-rich region of the irradiated
and annealed samples likely undergoes saturated trapping. Thus, the peak $S$ and $W$
values correspond directly to the defect rather than representing a mixture of bulk and
defect annihilation, ruling out the larger $2V_{\mathrm{As}}$--$2V_{\mathrm{Ga}}$ complex.
Furthermore, it has been reported that for low-resistivity GaAs or GaAs with impurity
concentrations below $10^{17}$ cm$^{-3}$, $V_{\mathrm{As}}$--$V_{\mathrm{Ga}}$ divacancies are
the dominant defect type and are stable up to 400$^\circ$C~\cite{dannefaer1986vacancy}, well above the
annealing temperatures used here.

\section*{Conclusions}
This work demonstrates that Ne$^{++}$ irradiation at 400 keV produces a sharply defined amorphous layer in GaAs, underlain by a crystalline region rich in dislocation loops and implanted gas. Over time and upon \textit{in situ} heating, recrystallization proceeds through two distinct regimes: slow, epitaxial regrowth along the [001] direction at lower temperatures, followed by accelerated twin-mediated recrystallization above 250\textdegree{}C. The process initiates at both the free surface and the crystalline interface but advances more rapidly from the top down. 4D-STEM order--disorder mapping reveals the full progression from amorphous to crystalline phases, while angular cross-correlation analyses confirm that, although long-range symmetry is partially restored, recrystallized regions retain signatures of lattice distortion and twin boundaries. Recrystallization restores overall orientation and long-range ordering but introduces local disorder, as evidenced by broadened CCF peaks and increased angular symmetry contributions.

Despite the recovery of structural order, spectroscopic characterization shows that the electronic properties of the recrystallized layer are degraded compared to undamaged GaAs. Low-loss EELS reveals plasmon damping and energy shifts indicative of high defect concentrations, while band edge features suggest the presence of localized states and electronic disorder. In contrast, the amorphous region, though structurally disordered, shows a plasmon response more similar to that of pristine GaAs. Positron DBS shows that, upon irradiation, GaAs develops a defect-rich near-surface region with extremely high vacancy concentrations. Annealed GaAs partially recrystallizes, but the defect-rich region persists and is dominated by stable $V_{\mathrm{As}}$--$V_{\mathrm{Ga}}$ divacancies. These findings highlight a critical disconnect between structural coherence and electronic functionality in ion-irradiated GaAs. Structural metrics alone do not capture the extent of damage or recovery; instead, full evaluation requires correlative analysis across structural, symmetry, and spectroscopic modalities. Taken together, these results underscore the importance of nanoscale characterization in understanding and controlling defect-driven processes in compound semiconductors.

\section*{Methods}

\subsection*{Sample preparation}
GaAs wafers with the [001] surface orientation were purchased from MTI Corporation. Each wafer was cleaved into 3 mm × 3 mm pieces for ion irradiation. Irradiations were carried out at the Ion Beam Materials Laboratory (IBML) at Los Alamos National Laboratory using a 200 kV Danfysik ion implanter. Samples were mounted on a nickel block using double-sided carbon tape to ensure good thermal contact, and active cooling was employed to maintain the sample temperature below 35\textdegree{}C. The target chamber pressure was maintained at approximately 1 × 10$^{-7}$ torr throughout the process. Ne$^{++}$ ions at 400 keV were used for irradiation, with a total fluence of 2 × 10$^{16}$ ions/cm$^2$. The resulting damage profile is shown in Figure~\ref{fig:overview}a.

Cross-sections for TEM analysis were prepared using a Helios G4 UX dual-beam SEM-FIB. Thinning was performed iteratively on both sides of the lamellae with decreasing acceleration voltages and currents. The sample was placed on a Protochips thermoelectric MEMS chip. Final thinning was performed with the lamella on the chip at 2 keV with a current of 86 pA to minimize material redeposition and damage, resulting in a thickness range of approximately 62 to 78 nm, based on low-loss EELS measurements.

After approximately four weeks at ambient temperature, partial recrystallization was observed in the top layer of the amorphous region and, to a lesser extent, in the region above the bulk GaAs containing dislocation loops, as shown in Figure~\ref{fig:overview}c.

\subsection*{STEM, TEM, and EDS imaging}

STEM images and energy-dispersive X-ray spectroscopy (EDS) maps were acquired using an image-corrected FEI Titan operated at 300 keV, equipped with a OneView 4K CCD camera. EDS was performed to verify the composition of the GaAs and to confirm that the distributions of Ga and As remained uniform following irradiation. The Ne concentration was found to have a maximum of approximately 1.2 at.\%, in reasonable agreement with SRIM simulations. Bright-field TEM imaging in Fig. S1 of the Supplemental Information was performed on the same FEI Titan using a 60 $\mu$m C2 aperture and an intensional defocus of roughly 200 nm to increase phase contrast and confirm that the amorphous layer extends from the GaAs surface immediately following irradiation.

\subsection*{4D-STEM data acquisition}

4D-STEM was acquired on an image-corrected FEI Titan operated at 300 keV with a OneView 4K CCD camera. Datasets were collected in microprobe mode with a camera length of 770 mm and probe size was defined by 50 $\mu$m aperture, giving a 0.76 $\mu$mrad convergence angle. The small convergence angle was selected to enhance the amorphous signal, creating a probe of 2.6 nm in real space. The data was acquired with a step size of 5 nm, resulting in partial overlap between adjacent probe positions. The dwell time was selected to increase signal without damaging the detector. 

\subsection*{Analysis of 4D-STEM data}
4D-STEM data were analyzed using the \texttt{py4DSTEM} Python package~\cite{savitzky2021py4dstem}, \texttt{scikit-learn}~\cite{scikitlearn}, and custom scripts. General data processing was carried out with \texttt{py4DSTEM}. Processing steps included fitting the beam center origin, correcting for any drift in the origin across the dataset, and performing a polar transform of all diffraction patterns. This transformation enabled fitting the amorphous ring as a background signal and allowed the amorphous-to-crystalline signal ratio to be determined for each pattern.

Bragg peaks arising from crystalline regions were then detected in each diffraction pattern. Using the ratio of Bragg peak intensity to the radially integrated amorphous background, the order--disorder maps shown in Fig.~\ref{fig:ordering} were generated. Colormap ranges were chosen to highlight subtle variations in the amorphous layer and to reveal the transition from ordered to disordered regions across crystalline--amorphous interfaces.

The angular symmetries in the electron nanodiffraction patterns were measured following the method described by Liu et al.~(2015)~\cite{liu2015interpretation}. Dynamical diffraction effects can distort the diffraction volume, complicating the interpretation of symmetry. To quantify angular symmetries, we computed the angular cross-correlation function (CCF) of the nanodiffraction patterns. The four-point CCF is defined as a function of the probe position $\mathbf{r} = (x, y)$ and the scattering vector $k$:

\begin{align*}
C(\mathbf{r},k,\Delta) = \frac{\langle I(k,\varphi)I(k,\varphi+\Delta) \rangle - \langle I(k,\varphi)\rangle^2}{\langle I(k,\varphi)\rangle^2}
\end{align*}
where $I(k,\varphi)$ is the diffracted intensity at a given scattering vector $k$, averaged over the azimuthal angle $\varphi$~\cite{liu2015interpretation}.

The CCFs shown in Fig.~\ref{fig:symmetry} were calculated using a narrow band of $k$-space corresponding to the region where the amorphous ring and the \{111\} Bragg peaks appear. This was done to selectively capture rotational correlations associated with the amorphous-to-crystalline transition in that specific portion of $k$-space (4.1 - 8.4 mrad). In contrast, the $n$-fold symmetry maps shown in Fig.~\ref{fig:symmetrymap} were computed using all of $k$-space beyond the central bright spot. The rationale for this approach is that the emergence of twins significantly enhances the two-fold symmetry CCF intensity. However, to probe crystallinity involving polycrystallinity and superimposed twins, higher-order symmetries—such as four-fold—must also be considered across the full $k$-space. For superimposed twins, streaking is observed in some of the diffraction patterns along both the [111] and [$\overline{1}$11] directions. This explains the observed increase in two-fold symmetry at 200\textdegree{}C within the amorphous region, as well as its subsequent reduction at the next temperature step when twins coalesce into grains. Modeling symmetry in amorphous materials is particularly challenging. To avoid confounding effects associated with $n$-fold symmetries that are divisible by odd numbers, such symmetries were excluded from the measurements shown in Fig.~\ref{fig:symmetrymap}.

To classify diffraction patterns, k-means++ clustering in \texttt{scikit-learn} was performed using 2 and 3 clusters, which explained approximately 75\% and 80\% of the variance, respectively. K-means++ initializes cluster centers by selecting the first center at random and then weighting the probability of subsequent centers by the squared distance to the nearest existing center. This approach reduces sensitivity to initialization, improves convergence, and typically yields more consistent clustering outcomes compared to random seeding. Clustering was applied to Region 1 data collected at 25\textdegree{}C and within the temperature range where rapid recrystallization was observed: 200\textdegree{}C, 250\textdegree{}C, 300\textdegree{}C. The physical motivation behind this approach was to detect either (1) features indicative of amorphous versus crystalline signal in low-temperature datasets, or (2) distinguish among amorphous, crystalline, and intermediate or unclassifiable signal in higher temperature datasets. This analysis was used to establish thresholds distinguishing amorphous from crystalline pixels. A random state was used for all clustering runs. The number of initializations, (\texttt{n-init}), was set to the default value of 10, and the maximum number of iterations, (\texttt{max-iter}), was set to 20; however, convergence was consistently achieved within 13 iterations. For clustering, diffraction patterns were standardized and binned by a factor of 2, resulting in a final resolution of $256 \times 256$ pixels, in order to improve signal-to-noise ratio and reduce computational cost. Beam/origin sway was correcter for prior to clustering.

\subsection*{Electron energy-loss spectroscopy}

Monochromated STEM-EELS was used to characterize the low-loss energy-loss features of irradiation-damaged GaAs in four distinct structural regions: the top recrystallized layer, the amorphous layer, the crystalline layer containing dense dislocation loops, and the undamaged bulk crystal. EELS measurements were conducted on the TEAM I microscope operating in scanning transmission electron microscopy (STEM) mode. TEAM I is equipped with a GIF Continuum detector and energy filtering capabilities. Wien-filter monochromation and an energy dispersion of 9 meV per channel, provided an energy resolution of 0.2 eV—sufficient to resolve features in the plasmon and valence energy regime up to 23 eV.

The third condenser lens current was adjusted to achieve a convergence angle of 20.5 mrad, producing a probe current of approximately 0.2 nA. EEL spectra were acquired using a Gatan K3 single-electron detector. EELS maps were collected from regions with thicknesses (in units of the inelastic mean free path) of $t/\lambda \leq 0.48$ ($t = 78$ nm) to minimize plural scattering due to sample thickness. Maps were acquired with a 5 nm step size and a 0.5\% frame/pixel live time. For each of the four structural regions, spectra were summed over an area 400 nm wide and 60 nm deep. To isolate the single-scattering distribution, residual plural scattering was removed by deconvolving the spectra with the zero-loss peak (ZLP), which was subsequently subtracted. Kramers–Krönig analysis was then applied to determine the energy-loss function using Gatan's DigitalMicrograph software~\cite{costantini2023}.

\subsection*{Positron annihilation spectroscopy}
Positron DBS data were collected using the variable energy positron beam at the Ion Beam
Materials Laboratory (IBML) at Los Alamos National Laboratory. Moderated positrons from a
$\sim 12$~mCi $^{22}$Na source are magnetically guided to a target chamber $\sim$5 m from the
source chamber. A closely related system has been described previously~\cite{jones2025positron},
with minor differences in the configuration of the transport stage and the target chamber, which
here is configured to incorporate two convergent ion beams for \emph{in situ} irradiation studies.
Spectra are collected using two Canberra BE3830P-DET HPGe detectors situated on opposing sides
of the target chamber in conjunction with two Lynx Digital Signal Analyzers. Due to presently
uncorrected beam drift resulting from time-varying stray magnetic fields from neighboring labs,
large sample wafers (1--2 inch square or rectangular forms) were selected to minimize the risk of
background contamination of recorded spectra. Additionally, data were collected in short duration
runs (10--30 min increments), corrected for amplifier drift post-collection, and then combined for
final analysis. Typical measurements at each energy required 1--3 hours of cumulative data
collection time. 

The $S$ and $W$ parameters are defined from the Doppler-broadened annihilation spectra as the
fraction of counts in the ranges 510.2--511.8 keV for $S$ and $511 \pm 1.7$ to $511 \pm 3.2$ keV for
$W$, relative to the total counts in the range 507.8--514.2 keV. Fitted bulk $S$ and $W$ values
($0.5551 \pm 0.0007$ and $0.0926 \pm 0.0002$, respectively) were used to normalize the
experimental curves. The $S/S_{\mathrm{bulk}}$ values are all normalized to the fitted $S_{\mathrm{bulk}}$ of the pristine sample. The diffusion length was determined by fitting the experimental $S(E)$ and
$W(E)$ profiles with a positron diffusion-trapping model. From the fitted diffusion length, the
concentration of vacancy-type defects was estimated by comparison with established models.  

Uncertainties in $S$ and $W$ were determined from counting statistics, and fits to experimental
data were constrained by the bulk parameters obtained from the undamaged sample. For annealed
samples, deviations from bulk values were interpreted as arising from positron trapping at specific
defect types, with saturation trapping assumed in regions of high defect concentration.
Comparisons with literature values of $S$/$W$ ratios for monovacancies and divacancies in GaAs
enabled defect assignment~\cite{borner1999large}.

\section*{Acknowledgments}

This work was primarily supported by the Laboratory Directed Research and Development (LDRD) program of Los Alamos National Laboratory under project number 20240879PRD4. This work was also supported by LDRD project number 20240033DR. The positron annihilation spectroscopy was supported by FUTURE (Fundamental Understanding of Transport Under Reactor Extremes), an Energy Frontier Research Center funded by the U.S. Department of Energy (DOE), Office of Science, Basic Energy Sciences (BES). This work was performed, in part, at the Center for Integrated Nanotechnologies, an Office of Science User Facility operated for the U.S. Department of Energy (DOE) Office of Science. Los Alamos National Laboratory is operated by Triad National Security, LLC, for the National Nuclear Security Administration of U.S. Department of Energy (Contract No. 89233218CNA000001). Work at the Molecular Foundry was supported by the Office of Science, Office of Basic Energy Sciences, of the U.S. Department of Energy under Contract No. DE-AC02- 05CH11231.

\section*{Data Availability}
All data needed to evaluate the conclusions in the paper are present in the paper and/or the Supplementary Materials. The raw 4D-STEM data are openly available at 10.5281/zenodo.17416428. Additional data are available upon reasonable request from the corresponding author.

\bibliography{main}% Produces the bibliography via BibTeX.

@PREAMBLE{
 "\providecommand{\noopsort}[1]{}" 
 # "\providecommand{\singleletter}[1]{#1}%" 
}

@article{kreller2019massively,
	title = {Massively enhanced ionic transport in irradiated crystalline pyrochlore},
	volume = {7},
	issn = {2050-7496},
	doi = {10.1039/C8TA10967B},
	number = {8},
	journal = {Journal of Materials Chemistry A},
	author = {Kreller, Cortney R. and Valdez, James A. and Holesinger, Terry G. and Morgan, Jonathan and Wang, Yongqiang and Tang, Ming and Garzon, Fernando H. and Mukundan, Rangachary and Brosha, Eric L. and Uberuaga, Blas P.},
	year = {2019},
	note = {Publisher: The Royal Society of Chemistry},
	pages = {3917--3923},
}

@article{janish2020insitu,
	title = {In-situ re-crystallization of heavily-irradiated {Gd2Ti2O7}},
	volume = {194},
	issn = {1359-6454},
	doi = {10.1016/j.actamat.2020.04.026},
	journal = {Acta Materialia},
	author = {Janish, Matthew T. and Schneider, Matthew M. and Valdez, James A. and McClellan, Kenneth J. and Byler, Darrin D. and Wang, Yongqiang and Chen, Di and Holesinger, Terry G. and Uberuaga, Blas P.},
	month = aug,
	year = {2020},
	pages = {403--411},
}

@article{zhang2025study,
	title = {Study of the recrystallization behaviors induced by annealing and irradiation on amorphous {SiC}},
	volume = {137},
	issn = {0021-8979},
	doi = {10.1063/5.0232414},
	number = {1},
	urldate = {2025-10-19},
	journal = {Journal of Applied Physics},
	author = {Zhang, Zijun and Jiang, Shengming and Hu, Xiaotian and Zhang, Jian},
	month = jan,
	year = {2025},
	pages = {015901},
}

@article{niu2023structural,
	title = {Structural and thermodynamic evolution of an amorphous {SiOC} ceramic after swift heavy ion irradiation},
	volume = {242},
	issn = {1359-6454},
	doi = {10.1016/j.actamat.2022.118475},
	journal = {Acta Materialia},
	author = {Niu, Min and Jayanthi, K. and Gao, Hongfei and Solomon, Alexandre P. and O'Quinn, Eric C. and Su, Lei and Qin, Yuanbin and Toimil-Molares, Maria Eugenia and Wang, Hongjie and Lang, Maik and Navrotsky, Alexandra},
	year = {2023},
	pages = {118475},
}

@article{pearton2021review,
	title = {Review—{Radiation} {Damage} in {Wide} and {Ultra}-{Wide} {Bandgap} {Semiconductors}},
	volume = {10},
	issn = {2162-8777},
	doi = {10.1149/2162-8777/abfc23},
	number = {5},
	journal = {ECS Journal of Solid State Science and Technology},
	author = {Pearton, S. J. and Aitkaliyeva, Assel and Xian, Minghan and Ren, Fan and Khachatrian, Ani and Ildefonso, Adrian and Islam, Zahabul and Jafar Rasel, Md Abu and Haque, Aman and Polyakov, A. Y. and Kim, Jihyun},
	month = may,
	year = {2021},
	note = {Publisher: IOP Publishing},
	pages = {055008},
}

@article{costantini2023,
	title = {Analysis of {Plasmon} {Loss} {Peaks} of {Oxides} and {Semiconductors} with the {Energy} {Loss} {Function}},
	volume = {16},
	issn = {1996-1944},
	doi = {10.3390/ma16247610},
	number = {24},
	urldate = {2024-05-20},
	journal = {Materials},
	author = {Costantini, Jean-Marc and Ribis, Joël},
	month = dec,
	year = {2023},
	pmid = {38138750},
	pmcid = {PMC10744408},
	pages = {7610},
}

@article{gammer2018local,
	title = {Local nanoscale strain mapping of a metallic glass during in situ testing},
	volume = {112},
	issn = {0003-6951},
	doi = {10.1063/1.5025686},
	number = {17},
	journal = {Applied Physics Letters},
	author = {Gammer, Christoph and Ophus, Colin and Pekin, Thomas C. and Eckert, Jürgen and Minor, Andrew M.},
	month = apr,
	year = {2018},
	pages = {171905},
}

@article{kennedy2025mapping,
	title = {Mapping strain and structural heterogeneities around bubbles in amorphous ionically conductive {Bi2O3}},
	volume = {256},
	issn = {0264-1275},
	doi = {10.1016/j.matdes.2025.114282},
	journal = {Materials \& Design},
	author = {Kennedy, Ellis Rae and Ribet, Stephanie M. and Winter, Ian S. and Kohnert, Caitlin A. and Wang, Yongqiang and Bustillo, Karen C. and Ophus, Colin and Derby, Benjamin K.},
	month = aug,
	year = {2025},
	pages = {114282},
}

@incollection{iles2004photovoltaic,
	address = {New York},
	title = {Photovoltaic {Conversion}: {Space} {Applications}},
	isbn = {978-0-12-176480-7},
	shorttitle = {Photovoltaic {Conversion}},
	booktitle = {Encyclopedia of {Energy}},
	publisher = {Elsevier},
	author = {Iles, Peter A.},
	editor = {Cleveland, Cutler J.},
	month = jan,
	year = {2004},
	doi = {10.1016/B0-12-176480-X/00332-6},
	pages = {25--33},
}

@article{yoo2024unsupervised,
	title = {Unsupervised machine learning and cepstral analysis with {4D}-{STEM} for characterizing complex microstructures of metallic alloys},
	volume = {10},
	copyright = {2024 The Author(s)},
	issn = {2057-3960},
	doi = {10.1038/s41524-024-01414-3},
	number = {1},
	journal = {npj Computational Materials},
	author = {Yoo, Timothy and Hershkovitz, Eitan and Yang, Yang and da Cruz Gallo, Flávia and Manuel, Michele V. and Kim, Honggyu},
	month = sep,
	year = {2024},
	note = {Publisher: Nature Publishing Group},
	pages = {223},
}

@article{kim2025operando,
	title = {Operando {Heating} and {Cooling} {Electrochemical} {4D}-{STEM} {Probing} {Nanoscale} {Dynamics} at {Solid}–{Liquid} {Interfaces}},
	volume = {147},
	issn = {0002-7863},
	doi = {10.1021/jacs.5c05005},
	number = {27},
	journal = {Journal of the American Chemical Society},
	author = {Kim, Sungin and Briega-Martos, Valentin and Liu, Shikai and Je, Kwanghwi and Shi, Chuqiao and Stephens, Katherine Marusak and Zeltmann, Steven E. and Zhang, Zhijing and Guzman-Soriano, Rafael and Li, Wenqi and Jiang, Jiahong and Choi, Juhyung and Negash, Yafet J. and Walden, Franklin S. II and Marthe, Nelson L. Jr. and Wellborn, Patrick S. and Guo, Yaofeng and Damiano, John and Han, Yimo and Thiede, Erik H. and Yang, Yao},
	month = jul,
	year = {2025},
	note = {Publisher: American Chemical Society},
	pages = {23654--23671},
}

@article{wochner2009xray,
	title = {X-ray cross correlation analysis uncovers hidden local symmetries in disordered matter},
	volume = {106},
	doi = {10.1073/pnas.0905337106},
	number = {28},
	journal = {Proceedings of the National Academy of Sciences},
	author = {Wochner, Peter and Gutt, Christian and Autenrieth, Tina and Demmer, Thomas and Bugaev, Volodymyr and Ortiz, Alejandro Díaz and Duri, Agnès and Zontone, Federico and Grübel, Gerhard and Dosch, Helmut},
	month = jul,
	year = {2009},
	note = {Publisher: Proceedings of the National Academy of Sciences},
	pages = {11511--11514},
}

@article{ophus2019four,
	title = {Four-{Dimensional} {Scanning} {Transmission} {Electron} {Microscopy} ({4D}-{STEM}): {From} {Scanning} {Nanodiffraction} to {Ptychography} and {Beyond}},
	volume = {25},
	issn = {1431-9276},
	doi = {10.1017/S1431927619000497},
	number = {3},
	journal = {Microscopy and Microanalysis},
	author = {Ophus, Colin},
	month = jun,
	year = {2019},
	pages = {563--582},
}

@article{hamadani2022visualizing,
	title = {Visualizing localized, radiative defects in {GaAs} solar cells},
	volume = {12},
	copyright = {2022 This is a U.S. Government work and not under copyright protection in the US; foreign copyright protection may apply},
	issn = {2045-2322},
	doi = {10.1038/s41598-022-19187-4},
	number = {1},
	journal = {Scientific Reports},
	author = {Hamadani, Behrang H. and Stevens, Margaret A. and Conrad, Brianna and Lumb, Matthew P. and Schmieder, Kenneth J.},
	month = sep,
	year = {2022},
	note = {Publisher: Nature Publishing Group},
	pages = {14838},
}

@article{williams1966determination,
	title = {Determination of {Deep} {Centers} in {Conducting} {Gallium} {Arsenide}},
	volume = {37},
	issn = {0021-8979},
	doi = {10.1063/1.1708872},
	number = {9},
	journal = {Journal of Applied Physics},
	author = {Williams, Richard},
	month = aug,
	year = {1966},
	pages = {3411--3416},
}

@article{soininen1992high,
	title = {High-temperature positron diffusion in {Si}, {GaAs}, and {Ge}},
	volume = {46},
	doi = {10.1103/PhysRevB.46.13104},
	number = {20},
	journal = {Physical Review B},
	author = {Soininen, E.},
	year = {1992},
	pages = {13104--13118},
}

@article{gebauer2000defect,
	title = {Defect identification in {GaAs} grown at low temperatures by positron annihilation},
	volume = {87},
	issn = {0021-8979},
	url = {https://doi.org/10.1063/1.373549},
	doi = {10.1063/1.373549},
	journal = {Journal of Applied Physics},
	author = {Gebauer, J. and Börner, F. and Krause-Rehberg, R. and Staab, T. E. M. and Bauer-Kugelmann, W. and Kögel, G. and Triftshäuser, W. and Specht, P. and Lutz, R. C. and Weber, E. R. and Luysberg, M.},
	month = jun,
	year = {2000},
	pages = {8368--8379},
}

@article{borner1999large,
	title = {Large-depth defect profiling in {GaAs} wafers after saw cutting},
	volume = {149},
	issn = {0169-4332},
	doi = {10.1016/S0169-4332(99)00192-0},
	number = {1},
	journal = {Applied Surface Science},
	author = {Börner, F and Eichler, S and Polity, A and Krause-Rehberg, R and Hammer, R and Jurisch, M},
	month = aug,
	year = {1999},
	pages = {151--158},
}

@article{dannefaer1986vacancy,
	title = {Vacancy interactions in {GaAs}},
	volume = {60},
	issn = {0021-8979},
	doi = {10.1063/1.337452},
	number = {2},
	journal = {Journal of Applied Physics},
	author = {Dannefaer, S. and Kerr, D.},
	month = jul,
	year = {1986},
	pages = {591--594},
}

@article{jones2025positron,
	title = {Positron stopping in multilayer materials},
	volume = {37},
	issn = {0953-8984},
	doi = {10.1088/1361-648X/adc062},
number = {18},
	journal = {Journal of Physics: Condensed Matter},
	author = {Jones, A C L and Chung, T and Selim, F A},
	month = mar,
	year = {2025},
	note = {Publisher: IOP Publishing},
	pages = {185702},
}

@article{vanveen1995vepfit,
	series = {Proceedings of the {Sixth} {International} {Workshop} on {Slow}-{Positron} {Beam} {Techniques} for {Solids} and {Surfaces}},
	title = {{VEPFIT} applied to depth profiling problems},
	volume = {85},
	issn = {0169-4332},
	doi = {10.1016/0169-4332(94)00334-3},
	journal = {Applied Surface Science},
	author = {van Veen, A. and Schut, H. and Clement, M. and de Nijs, J. M. M. and Kruseman, A. and IJpma, M. R.},
	month = jan,
	year = {1995},
	pages = {216--224},
}

@article{corbel1988positron,
	title = {Positron-annihilation spectroscopy of native vacancies in as-grown {GaAs}},
	volume = {38},
	doi = {10.1103/PhysRevB.38.8192},
	number = {12},
	journal = {Physical Review B},
	author = {Corbel, C.},
	year = {1988},
	pages = {8192--8208},
}

@article{schulz1996optimum,
	title = {Optimum temperature for ion beam induced crystallization of {GaAs}},
	volume = {117},
	issn = {0168-583X},
	doi = {10.1016/0168-583X(96)00232-7},
	number = {1},
	urldate = {2025-10-22},
	journal = {Nuclear Instruments and Methods in Physics Research Section B: Beam Interactions with Materials and Atoms},
	author = {Schulz, R. and Bachmann, T. and Kaiser, U. and Glaser, E.},
	month = aug,
	year = {1996},
	pages = {207--209},
}

@article{herold1989investigation,
	title = {Investigation of {Low}-{Temperature} {Epitaxial} {Regrowth} of {Ion}-{Implanted} {Amorphous} {GaAs}},
	volume = {111},
	issn = {1521-396X},
	doi = {10.1002/pssa.2211110106},
	number = {1},
	journal = {physica status solidi (a)},
	author = {Herold, J. and Bartsch, H. and Wesch, W. and Götz, G.},
	year = {1989},
	note = {\_eprint: https://onlinelibrary.wiley.com/doi/pdf/10.1002/pssa.2211110106},
	pages = {59--70},
}

@article{selim2021positron,
	title = {Positron annihilation spectroscopy of defects in nuclear and irradiated materials- a review},
	volume = {174},
	issn = {1044-5803},
	doi = {10.1016/j.matchar.2021.110952},
	journal = {Materials Characterization},
	author = {Selim, F. A.},
	month = apr,
	year = {2021},
	pages = {110952},
}

@article{desnica-frankovic1999different,
	title = {Different recrystallization patterns of {Si}+ implanted {GaAs}},
	volume = {85},
	issn = {0021-8979},
	doi = {10.1063/1.370559},
	number = {11},
	journal = {Journal of Applied Physics},
	author = {Desnica-Franković, I. D.},
	month = jun,
	year = {1999},
	pages = {7587--7596},
}

@article{allen1960gallium,
	title = {Gallium {Arsenide} as a {Semi}-insulator},
	volume = {187},
	copyright = {1960 Springer Nature Limited},
	issn = {1476-4687},
	doi = {10.1038/187403b0},
	number = {4735},
	journal = {Nature},
	author = {Allen, J. W.},
	month = jul,
	year = {1960},
	note = {Publisher: Nature Publishing Group},
	pages = {403--405},
}

@article{smith2014facile,
	title = {Facile {Photochemical} {Preparation} of {Amorphous} {Iridium} {Oxide} {Films} for {Water} {Oxidation} {Catalysis}},
	volume = {26},
	issn = {0897-4756},
	doi = {10.1021/cm4041715},
	number = {4},
	journal = {Chemistry of Materials},
	author = {Smith, Rodney D. L. and Sporinova, Barbora and Fagan, Randal D. and Trudel, Simon and Berlinguette, Curtis P.},
	month = feb,
	year = {2014},
	note = {Publisher: American Chemical Society},
	pages = {1654--1659},
}

@article{zhang2021sodium,
	title = {Sodium-{Decorated} {Amorphous}/{Crystalline} {RuO2} with {Rich} {Oxygen} {Vacancies}: {A} {Robust} {pH}-{Universal} {Oxygen} {Evolution} {Electrocatalyst}},
	volume = {60},
	issn = {1521-3773},
	shorttitle = {Sodium-{Decorated} {Amorphous}/{Crystalline} {RuO2} with {Rich} {Oxygen} {Vacancies}},
	doi = {10.1002/anie.202106631},
	number = {34},
	journal = {Angewandte Chemie International Edition},
	author = {Zhang, Lijie and Jang, Haeseong and Liu, Huihui and Kim, Min Gyu and Yang, Dongjiang and Liu, Shangguo and Liu, Xien and Cho, Jaephil},
	year = {2021},
	pages = {18821--18829},
}

@article{jung2019amorphous,
	title = {Amorphous {FeZr} metal for multi-functional sensor in electronic skin},
	volume = {3},
	copyright = {2019 The Author(s)},
	issn = {2397-4621},
	doi = {10.1038/s41528-019-0051-7},
	number = {1},
	journal = {npj Flexible Electronics},
	author = {Jung, Minhyun and Lee, Eunha and Kim, Dongseuk and Kim, Kyungkwan and Yun, Changjin and Lee, Hyangsook and Kim, Heegoo and Rhie, Kungwon and Jeon, Sanghun},
	month = apr,
	year = {2019},
	note = {Publisher: Nature Publishing Group},
	pages = {8},
}

@article{kamiya2010material,
	title = {Material characteristics and applications of transparent amorphous oxide semiconductors},
	volume = {2},
	copyright = {2010 Tokyo Institute of Technology},
	issn = {1884-4057},
	doi = {10.1038/asiamat.2010.5},
	number = {1},
	journal = {NPG Asia Materials},
	author = {Kamiya, Toshio and Hosono, Hideo},
	month = jan,
	year = {2010},
	note = {Publisher: Nature Publishing Group},
	pages = {15--22},
}

@article{zhang2025unveiling,
	title = {Unveiling the microscopic origin of anomalous thermal conductivity in amorphous carbon},
	volume = {11},
	doi = {10.1126/sciadv.adx5007},
	number = {23},
	urldate = {2025-08-31},
	journal = {Science Advances},
	author = {Zhang, ZhongTing and Luo, Jian and Wu, HengAn and Ma, Hao and Zhu, YinBo},
	month = jun,
	year = {2025},
	note = {Publisher: American Association for the Advancement of Science},
	pages = {eadx5007},
}

@article{zhang2023anomalous,
	title = {Anomalous topological waves in strongly amorphous scattering networks},
	volume = {9},
	doi = {10.1126/sciadv.adg3186},
	number = {12},
	urldate = {2025-08-31},
	journal = {Science Advances},
	author = {Zhang, Zhe and Delplace, Pierre and Fleury, Romain},
	month = mar,
	year = {2023},
	note = {Publisher: American Association for the Advancement of Science},
	pages = {eadg3186},
}

@article{johnston2013radiation,
	title = {Radiation {Effects} in {Optoelectronic} {Devices}},
	volume = {60},
	issn = {1558-1578},
	doi = {10.1109/TNS.2013.2259504},
	number = {3},
	urldate = {2025-08-31},
	journal = {IEEE Transactions on Nuclear Science},
	author = {Johnston, Allan H.},
	month = jun,
	year = {2013},
	pages = {2054--2073},
	}

@article{srour1988radiation,
	title = {Radiation effects on microelectronics in space},
	volume = {76},
	issn = {1558-2256},
	doi = {10.1109/5.90114},
	number = {11},
	urldate = {2025-08-31},
	journal = {Proceedings of the IEEE},
	author = {Srour, J.R. and McGarrity, J.M.},
	month = nov,
	year = {1988},
	pages = {1443--1469},
}

@article{sorensen2020revealing,
	title = {Revealing hidden medium-range order in amorphous materials using topological data analysis},
	volume = {6},
	doi = {10.1126/sciadv.abc2320},
	number = {37},
	journal = {Science Advances},
	author = {Sørensen, Søren S. and Biscio, Christophe A. N. and Bauchy, Mathieu and Fajstrup, Lisbeth and Smedskjaer, Morten M.},
	month = sep,
	year = {2020},
	note = {Publisher: American Association for the Advancement of Science},
	pages = {eabc2320},
}

@article{nishio2013universal,
	title = {Universal {Medium}-{Range} {Order} of {Amorphous} {Metal} {Oxides}},
	volume = {111},
	doi = {10.1103/PhysRevLett.111.155502},
	number = {15},
	journal = {Physical Review Letters},
	author = {Nishio, Kengo and Miyazaki, Takehide and Nakamura, Hisao},
	month = oct,
	year = {2013},
	note = {Publisher: American Physical Society},
	pages = {155502},
}

@article{voyles2001structure,
	title = {Structure and physical properties of paracrystalline atomistic models of amorphous silicon},
	volume = {90},
	issn = {0021-8979},
	doi = {10.1063/1.1407319},
	number = {9},
	journal = {Journal of Applied Physics},
	author = {Voyles, P. M. and Zotov, N. and Nakhmanson, S. M. and Drabold, D. A. and Gibson, J. M. and Treacy, M. M. J. and Keblinski, P.},
	month = nov,
	year = {2001},
	pages = {4437--4451}
}

@article{ciocys2024establishing,
	title = {Establishing coherent momentum-space electronic states in locally ordered materials},
	volume = {15},
	copyright = {2024 The Author(s)},
	issn = {2041-1723},	doi = {10.1038/s41467-024-51953-y},
	journal = {Nature Communications},
	author = {Ciocys, Samuel T. and Marsal, Quentin and Corbae, Paul and Varjas, Daniel and Kennedy, Ellis and Scott, Mary and Hellman, Frances and Grushin, Adolfo G. and Lanzara, Alessandra},
	month = sep,
	year = {2024},
	note = {Publisher: Nature Publishing Group},
	pages = {8141},
}

@article{divincenzo1984long,
	title = {Long-range structural and electronic coherence in amorphous semiconductors},
	volume = {29},
	doi = {10.1103/PhysRevB.29.5934},
	number = {10},
	urldate = {2025-08-31},
	journal = {Physical Review B},
	author = {DiVincenzo, D. P. and Mosseri, R. and Brodsky, M. H. and Sadoc, J. F.},
	month = may,
	year = {1984},
	note = {Publisher: American Physical Society},
	pages = {5934--5936},
}

@inproceedings{brodsky1985structural,
	address = {New York, NY},
	title = {A {Structural} {Basis} for {Electronic} {Coherence} in {Amorphous} {Si} and {Ge}},
	isbn = {978-1-4615-7682-2},
	doi = {10.1007/978-1-4615-7682-2_178},
	booktitle = {Proceedings of the 17th {International} {Conference} on the {Physics} of {Semiconductors}},
	publisher = {Springer},
	author = {Brodsky, M. H. and DiVincenzo, D. P. and Mosseri, R. and Sadoc, J. F.},
	editor = {Chadi, James D. and Harrison, Walter A.},
	year = {1985},
	pages = {803--806},
}

@article{velisa2025revealing,
	title = {Revealing a {Pathway} for {Low}-{Temperature} {Recrystallization} in {Germanium}},
	volume = {n/a},
    year ={2025},
	issn = {2198-3844},
	doi = {10.1002/advs.202507630},
	number = {n/a},
	journal = {Advanced Science},
	author = {Velişa, Gihan and Zarkadoula, Eva and Iancu, Decebal and Mihai, Maria D. and Boulle, Alexandre and Tong, Yang and Chen, Da and Zhang, Yanwen and Weber, William J.},
	pages = {e07630},
}

@article{scikitlearn,
    title={Scikit-learn: Machine Learning in {P}ython},
    author={Pedregosa, F. and Varoquaux, G. and Gramfort, A. and Michel, V. and Thirion, B. and Grisel, O. and Blondel, M. and Prettenhofer, P. and Weiss, R. and Dubourg, V. and Vanderplas, J. and Passos, A. and Cournapeau, D. and Brucher, M. and Perrot, M. and Duchesnay, E.},
    journal={Journal of Machine Learning Research},
    volume={12},
    pages={2825--2830},
    year={2011},
   }

@article{savitzky2021py4dstem,
  title={py4{DSTEM}: A software package for four-dimensional scanning transmission electron microscopy data analysis},
  author={Savitzky, Benjamin H and Zeltmann, Steven E and Hughes, Lauren A and Brown, Hamish G and Zhao, Shiteng and Pelz, Philipp M and Pekin, Thomas C and Barnard, Edward S and Donohue, Jennifer and DaCosta, Luis Rangel and others},
  journal={Microscopy and Microanalysis},
  volume={27},
  number={4},
  pages={712--743},
  year={2021},
  publisher={Cambridge University Press}
}

@article{treacy2012local,
	title = {The {Local} {Structure} of {Amorphous} {Silicon}},
	volume = {335},
	doi = {10.1126/science.1214780},
	number = {6071},
	journal = {Science},
	author = {Treacy, M. M. J. and Borisenko, K. B.},
	month = feb,
	year = {2012},
	note = {Publisher: American Association for the Advancement of Science},
	pages = {950--953},
}

@article{rosset2025signatures,
	title = {Signatures of paracrystallinity in amorphous silicon from machine-learning-driven molecular dynamics},
	volume = {16},
	copyright = {2025 The Author(s)},
	issn = {2041-1723},
	doi = {10.1038/s41467-025-57406-4},
	number = {1},
	urldate = {2025-08-30},
	journal = {Nature Communications},
	author = {Rosset, Louise A. M. and Drabold, David A. and Deringer, Volker L.},
	month = mar,
	year = {2025},
	note = {Publisher: Nature Publishing Group},
	pages = {2360},
}

@article{nakayama2025symmetry,
	title = {Symmetry breaking of paracrystalline topology in amorphous silicon},
	volume = {15},
	issn = {2045-2322},
	doi = {10.1038/s41598-025-15737-8},
	number = {1},
	journal = {Scientific Reports},
	author = {Nakayama, Koji S. and Nishijima, Masahiko and Zhang, Yicheng and Inoue, Koji and Chen, Chuantong and Ueshima, Minoru and Suganuma, Katsuaki},
	month = aug,
	year = {2025},
	pages = {31539},
}

@article{singh2023impact,
	title = {Impact of radiation-induced point defects on thermal carrier decay processes in {GaAs}},
	volume = {242},
	issn = {1359-6454},
	doi = {10.1016/j.actamat.2022.118480},
	journal = {Acta Materialia},
	author = {Singh, Christopher N. and Uberuaga, Blas Pedro and Tobin, Stephen J. and Liu, Xiang-Yang},
	month = jan,
	year = {2023},
	pages = {118480},
}

@article{shuk1997,
	title = {Electronic {Properties} of {Bi2O3} {Based} {Solid} {Electrolytes}},
	volume = {623},
	copyright = {Copyright © 1997 Verlag GmbH \& Co. KGaA, Weinheim},
	issn = {1521-3749},
	doi = {10.1002/zaac.199762301140},
	number = {1-6},
	journal = {Zeitschrift für anorganische und allgemeine Chemie},
	author = {Shuk, P. and Wiemhöfer, H.-D. and Göpel, W.},
	year = {1997},
	pages = {892--896},
}

@incollection{batson2004electron,
	title = {Electron {Energy} {Loss} {Studies} in {Semiconductors}},
	copyright = {Copyright © 2004 Wiley-VCH Verlag GmbH \& Co. KGaA},
	isbn = {978-3-527-60549-1},
	booktitle = {Transmission {Electron} {Energy} {Loss} {Spectrometry} in {Materials} {Science} and {The} {EELS} {Atlas}},
	publisher = {John Wiley \& Sons, Ltd},
	author = {Batson, Philip E.},
	year = {2004},
	doi = {10.1002/3527605495.ch10},
	pages = {353--384},
}

@article{foerster2019plasmon,
	title = {Plasmon damping depends on the chemical nature of the nanoparticle interface},
	volume = {5},
	doi = {10.1126/sciadv.aav0704},
	number = {3},
	journal = {Science Advances},
	author = {Foerster, Benjamin and Spata, Vincent A. and Carter, Emily A. and Sönnichsen, Carsten and Link, Stephan},
	month = mar,
	year = {2019},
	pages = {eaav0704},
}

@article{aguillon2021plasmons,
	title = {Plasmons in {Graphene} {Nanostructures} with {Point} {Defects} and {Impurities}},
	volume = {125},
	issn = {1932-7447},
	doi = {10.1021/acs.jpcc.1c05182},
	number = {39},
	journal = {The Journal of Physical Chemistry C},
	author = {Aguillon, François and Marinica, Dana Codruta and Borisov, Andrei G.},
	month = oct,
	year = {2021},
	note = {Publisher: American Chemical Society},
	pages = {21503--21510},
}

@article{li2023quantitative,
	title = {A quantitative method to determine the region not influenced by injected interstitial and surface effects during void swelling in ion-irradiated metals},
	volume = {573},
	issn = {0022-3115},
	doi = {10.1016/j.jnucmat.2022.154140},
	journal = {Journal of Nuclear Materials},
	author = {Li, Yongchang and French, Aaron and Hu, Zhihan and Gabriel, Adam and Hawkins, Laura R. and Garner, Frank A. and Shao, Lin},
	month = jan,
	year = {2023},
	pages = {154140},
}

@article{shao2014effect,
	title = {Effect of defect imbalance on void swelling distributions produced in pure iron irradiated with 3.5 {MeV} self-ions},
	volume = {453},
	issn = {0022-3115},
	doi = {10.1016/j.jnucmat.2014.06.002},
	number = {1},
	journal = {Journal of Nuclear Materials},
	author = {Shao, Lin and Wei, C. -C. and Gigax, J. and Aitkaliyeva, A. and Chen, D. and Sencer, B. H. and Garner, F. A.},
	month = oct,
	year = {2014},
	pages = {176--181},
}

@article{kennedy2024insights,
	title = {Insights into defect kinetics, mass transport, and electronic structure from spectrum effects in ion-irradiated $\mathrm{Bi_2O_3}$},
	volume = {12},
	issn = {2050-7496},
	doi = {10.1039/D4TA05283H},
	pages = {31445--31458},
	number = {45},
	journal = {Journal of Materials Chemistry A},
	shortjournal = {J. Mater. Chem. A},
	author = {Kennedy, Ellis Rae and Valdez, James A. and Wang, Yongqiang and Ribet, Stephanie M. and Sickafus, Kurt E. and Kreller, Cortney R. and Uberuaga, Blas Pedro and Derby, Benjamin K.},
    year = {2024},
	note = {Publisher: The Royal Society of Chemistry},
}

@article{liu2025review,
	title = {Review of radiation-induced defects in {GaAs}},
	volume = {138},
	issn = {0021-8979},
	doi = {10.1063/5.0279267},
	number = {7},
	journal = {Journal of Applied Physics},
	author = {Liu, X.-Y. and Singh, C. N. and Kennedy, E. R. and Huang, S. and Fluckey, S. P. and Matthews, C. and Uberuaga, B. P.},
	month = aug,
	year = {2025},
	pages = {070701},
}

@article{ridley2024amorphous,
	title = {Amorphous and nanocrystalline halide solid electrolytes with enhanced sodium-ion conductivity},
	volume = {7},
	issn = {2590-2385},
	doi = {10.1016/j.matt.2023.12.028},
	number = {2},
	journal = {Matter},
	author = {Ridley, Phillip and Nguyen, Long Hoang Bao and Sebti, Elias and Han, Bing and Duong, George and Chen, Yu-Ting and Sayahpour, Baharak and Cronk, Ashley and Deysher, Grayson and Ham, So-Yeon and Oh, Jin An Sam and Wu, Erik A. and Tan, Darren H. S. and Doux, Jean-Marie and Clément, Raphaële and Jang, Jihyun and Meng, Ying Shirley},
	month = feb,
	year = {2024},
	pages = {485--499},
}

@incollection{kim2012uranium,
	address = {Oxford},
	title = {3.14 - {Uranium} {Intermetallic} {Fuels} ({U}–{Al}, {U}–{Si}, {U}–{Mo})},
	isbn = {978-0-08-056033-5},
	booktitle = {Comprehensive {Nuclear} {Materials}},
	publisher = {Elsevier},
	author = {Kim, Yeon Soo},
	editor = {Konings, Rudy J. M.},
	month = jan,
	year = {2012},
	doi = {10.1016/B978-0-08-056033-5.00112-9},
	pages = {391--422},
}

@article{liu2015interpretation,
	title = {Interpretation of angular symmetries in electron nanodiffraction patterns from thin amorphous specimens},
	volume = {71},
	issn = {2053-2733},
	doi = {10.1107/S2053273315011845},
	pages = {473--482},
	number = {5},
	journal = {Acta Crystallographica Section A: Foundations and Advances},
	shortjournal = {Acta Cryst A},
	author = {Liu, A. C. Y. and Lumpkin, G. R. and Petersen, T. C. and Etheridge, J. and Bourgeois, L.},
	date = {2015-09-01},
	langid = {english},
    year = {2015},
	note = {Publisher: International Union of Crystallography},
}

@article{kennedy2024exploring,
	title = {Exploring Structural Anisotropy in Amorphous {T}b-{C}o via Changes in Medium-Range Ordering},
	issn = {1431-9276},
	doi = {10.1093/mam/ozae113},
	pages = {ozae113},
	journal = {Microscopy and Microanalysis},
	shortjournal = {Microscopy and Microanalysis},
	author = {Kennedy, Ellis and Hollingworth, Emily and Ceballos, Alejandro and O’Mahoney, Daisy and Ophus, Colin and Hellman, Frances and Scott, Mary},
	date = {2024-11-14},
    year = {2024},
}

\end{document}

\vspace*{-1.5cm} % move up the title if needed
\begin{center}
    \Large
    \textbf{Supplemental Information for}\\[0.5em]
    ``Critical Disconnect Between Structural and Electronic Recovery in Amorphous GaAs during Recrystallization''
\end{center}
\vspace{0.5cm}

\begin{figure*}[!ht]
\centering
\includegraphics[width=0.9\textwidth]{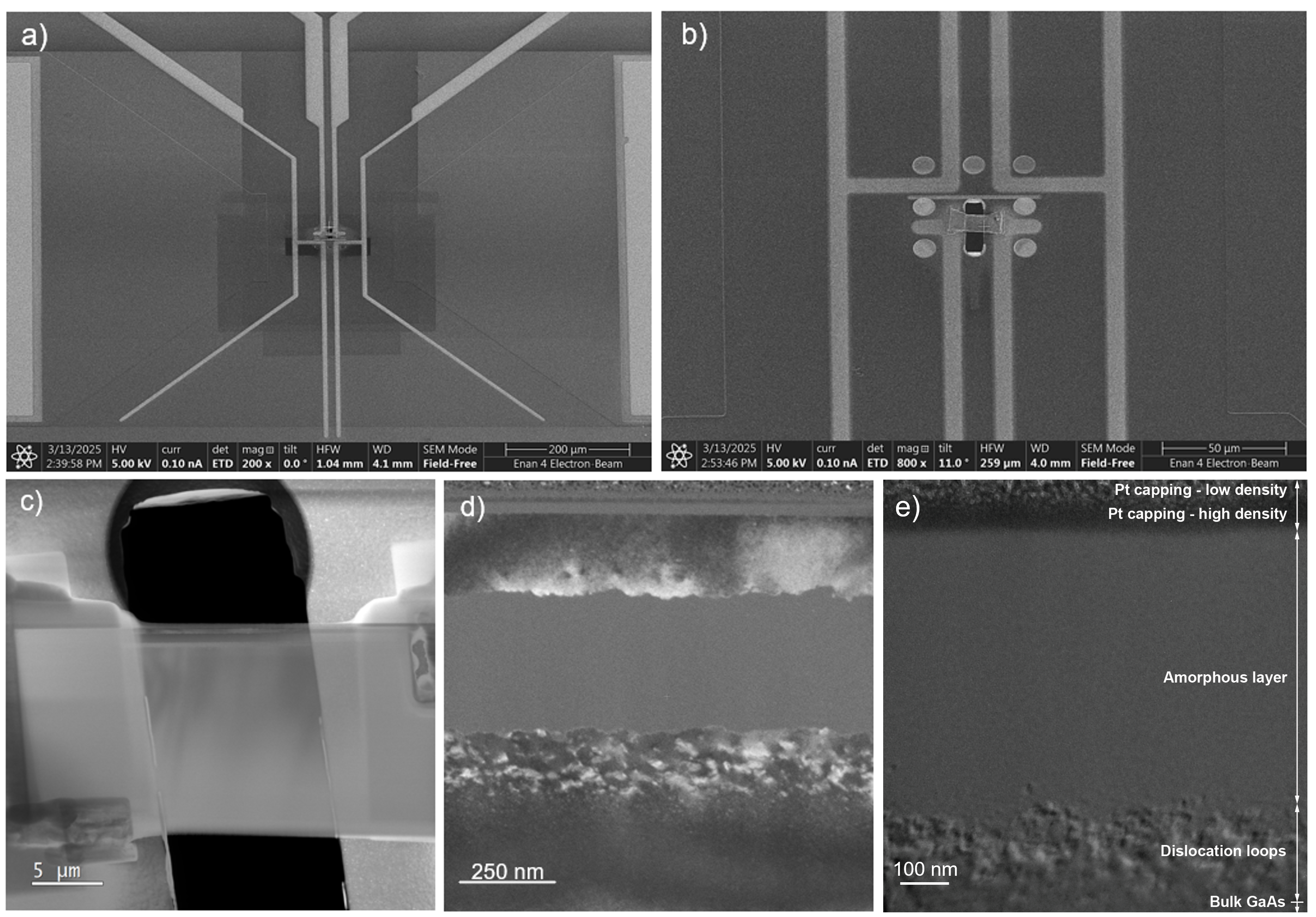}
\caption{\label{fig:figs1} (a,b) The FIB-prepared lamella was mounted onto a Protochips thermoelectric MEMS chip for \textit{in situ} heating. A final low-energy cleaning step was performed at 2 keV to minimize surface damage. (c) The lamella is positioned over a hole in the support grid, with the Pt capping layer and ion-damaged regions clearly visible above the underlying pristine GaAs. (d) SEM-STEM image showing the layered structure: recrystallized region at the top, followed by the amorphous layer, a band of dislocation loops, and undamaged bulk GaAs. (e) TEM image acquired immediately following Ne$^{++}$ irradiation. A 60 $\mu$m C2 aperture and intentional defocus were applied to capture phase contrast from structural variations. No recrystallization is observed at the GaAs surface beneath the Pt layer. Pt was applied to the GaAs surface in the FIB, first at low current (gentle, higher density) to avoid altering the surface, then at a higher current (lower density).}
\end{figure*}

\begin{figure*}[!ht]
\centering
\includegraphics[width=0.80\textwidth]{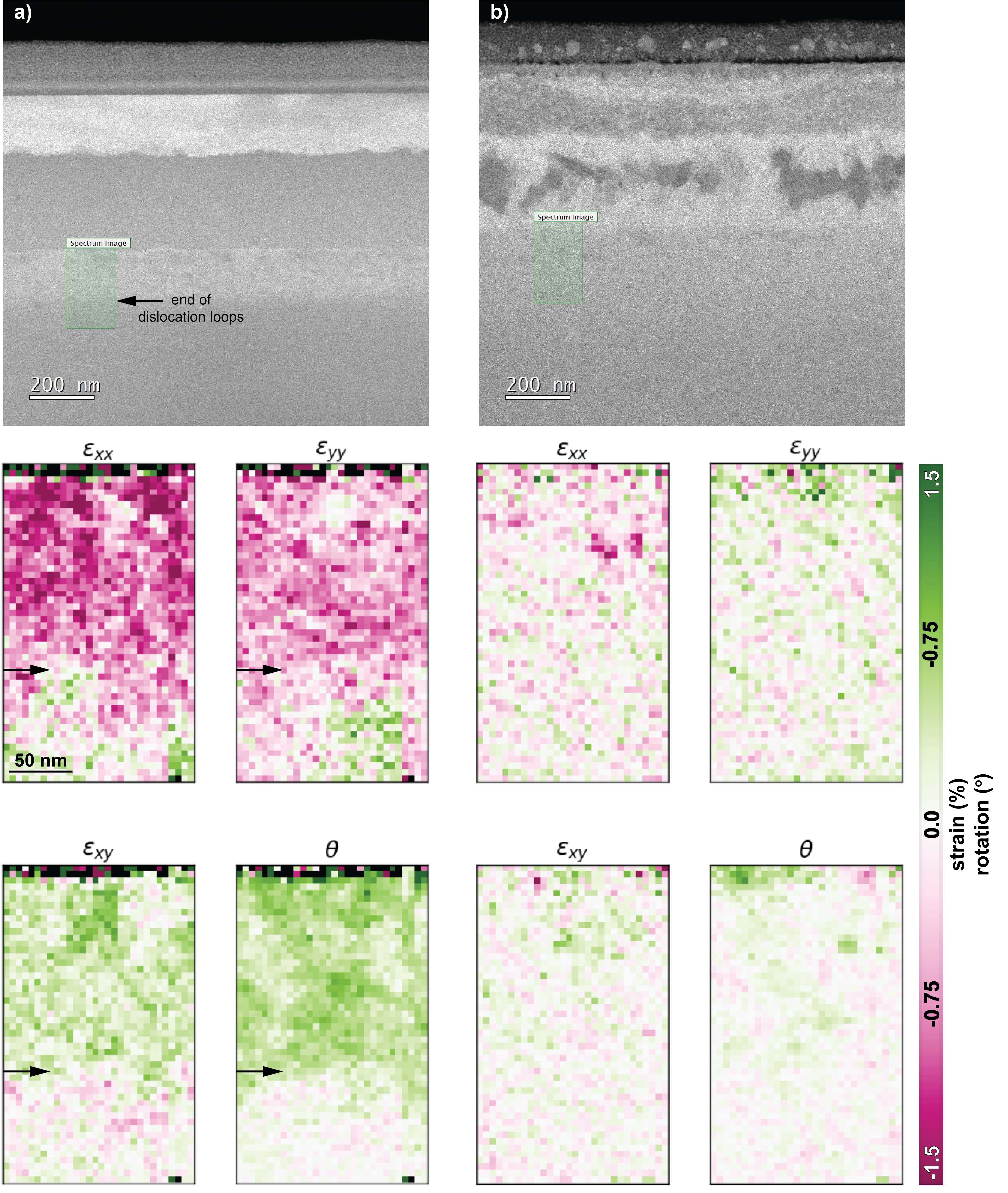}
\caption{\label{fig:figs4} Additional 4D-STEM datasets were collected across the dislocation loop region (a) prior to heating to 400\textdegree{}C and (b) after heating to 400\textdegree{}C followed by cooling to room temperature, to assess the effects of thermal treatment on the loop-induced damage. 4D-STEM scan regions are indicated by green boxes. Arrows indicate the base of the dislocation loop layer; material below the arrows is undamaged GaAs. (a) Before heating, the dislocation loops and associated radiation damage result in a region of compressive $\varepsilon_{xx}$ and $\varepsilon_{yy}$ strain (up to ~1.5\%) directly beneath the amorphous layer. (b) After heating, the strain is significantly reduced, and the former loop region becomes indistinguishable from the bulk in the strain maps. This relaxation likely results from thermally activated atomic mobility and defect recombination at elevated temperature. The undamaged bulk was used as the reference zero-strain state. While a more detailed analysis is ongoing, this result suggests that annealing at 400\,\textdegree{}C removes dislocation-loop-associated strain, and the loops themselves are no longer visible in the STEM image.}
\end{figure*}

\begin{figure*}[!ht]
\centering
\includegraphics[width=0.95\textwidth]{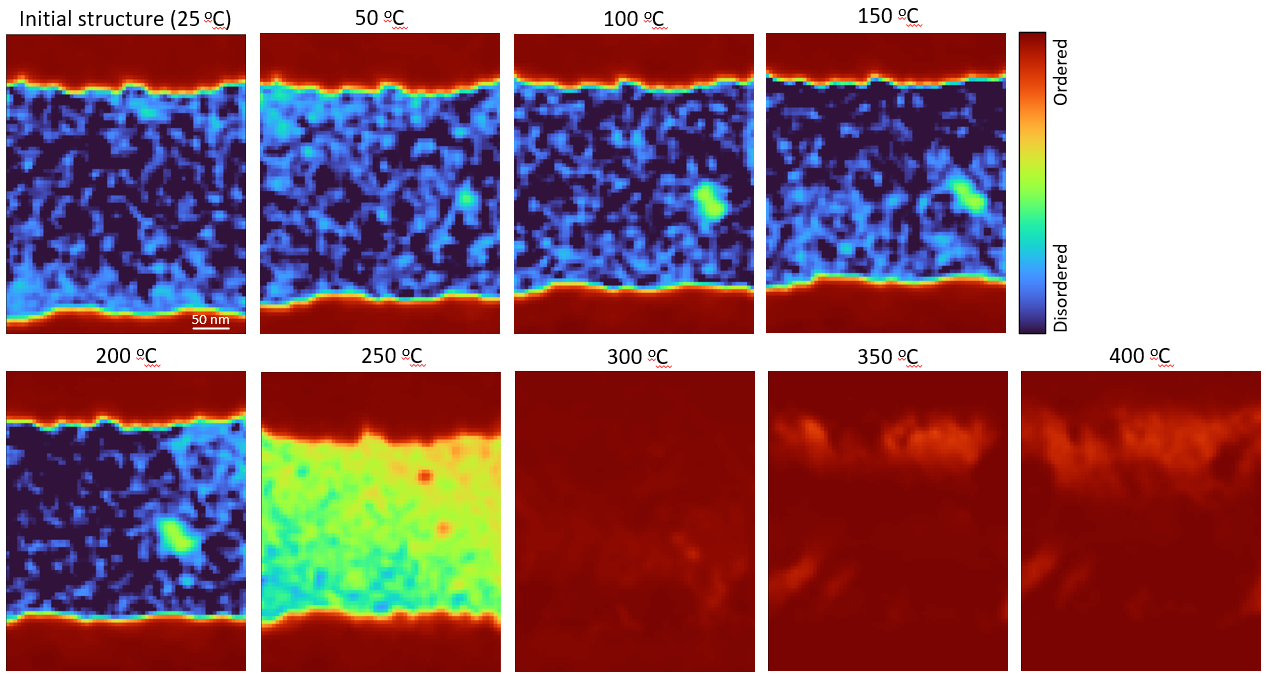}
\caption{\label{fig:figs2} Order-disorder maps from region 1 covering all temperature steps from 25\textdegree{}C to 400\textdegree{}C in 50\textdegree{C} increments. The spot of higher relative ordering within the amorphous layer on the right side of the region is a fiducial marked created by holding the beam in place. Recrystallization is complete by 300\textdegree{}C. }
\end{figure*}

\pagestyle{empty}

\begin{figure*}[!ht]
\centering
\includegraphics[width=0.9\textwidth]{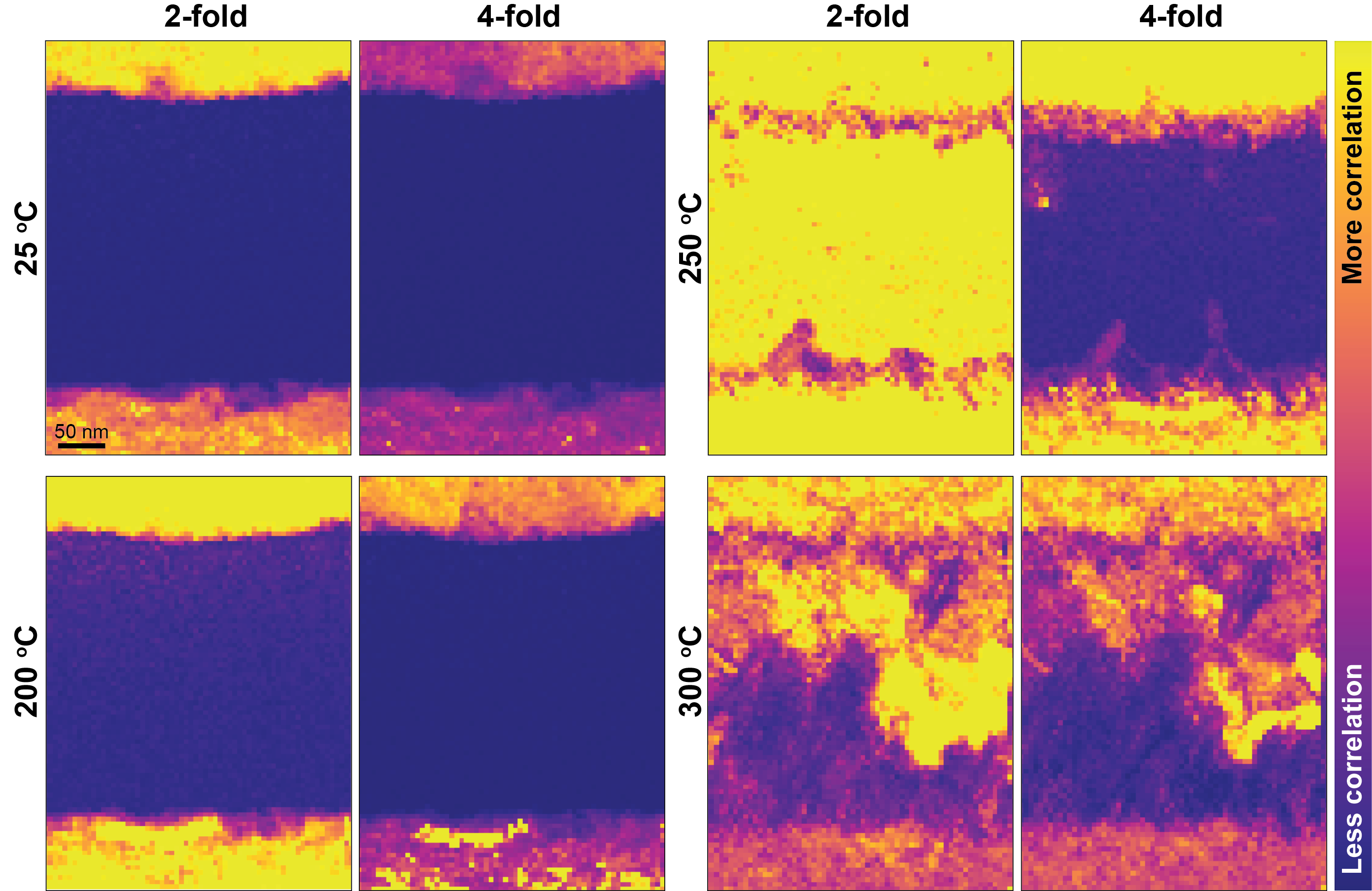}
\caption{\label{fig:figs3} As with region 1, \textit{n}-fold angular correlation maps were generated from 4D-STEM datasets of region 2 by measuring periodicity in each diffraction pattern. At 200\,\textdegree{}C, the 2-fold map shows initial signs of top-down recrystallization. By 250\,\textdegree{}C, the formation of twins extending from the crystalline layers into the amorphous matrix is clearly evident.}
\end{figure*}

\begin{figure*}[!ht]
\centering
\includegraphics[width=0.90\textwidth]{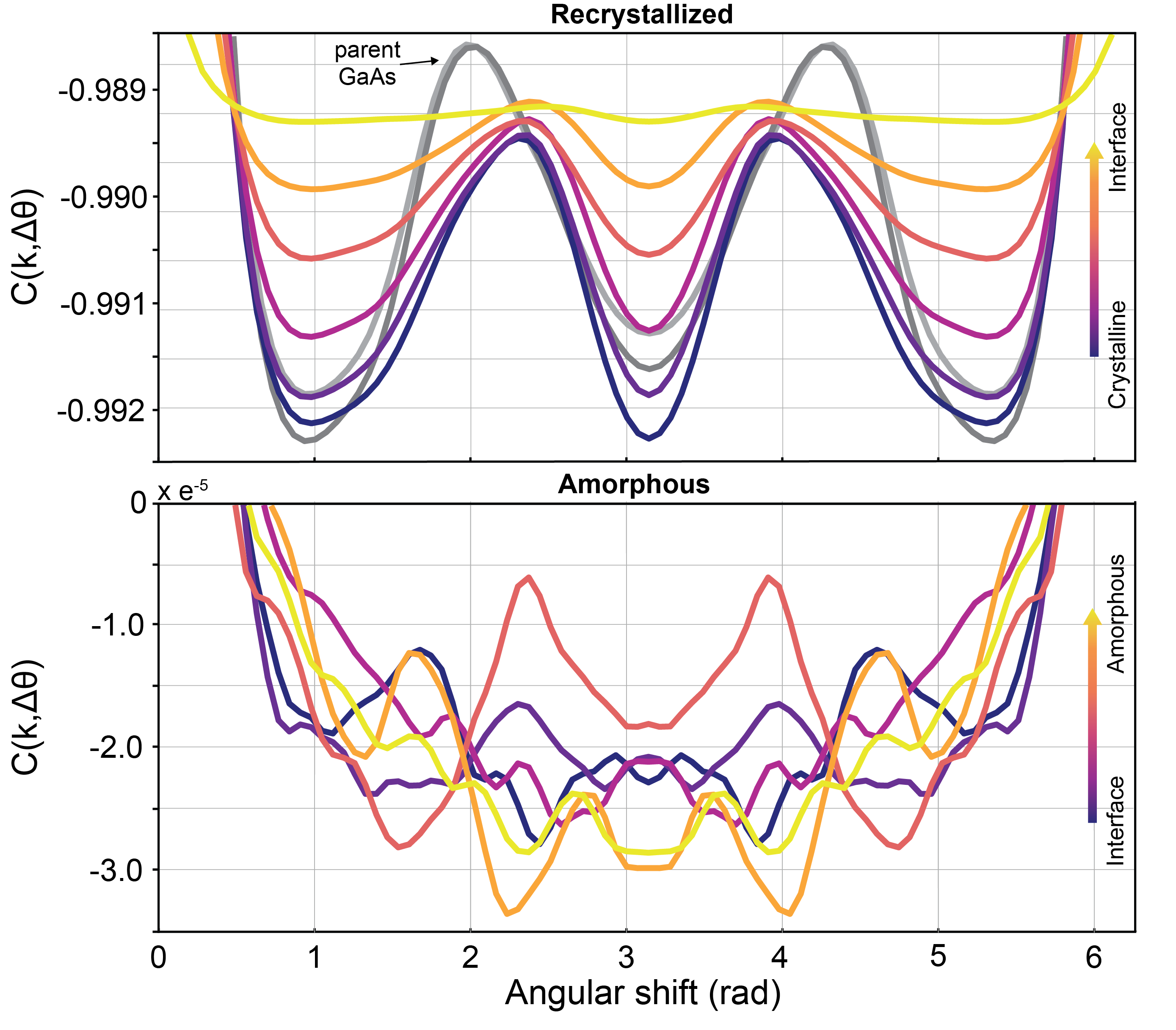}
\caption{\label{fig:figs6} Additional CCF plots collected from the recrystallized and amorphous regions. CCFs from the recrystallized layer were acquired at increasing proximity to the amorphous interface to assess how angular correlations evolve near the disordered boundary. Peak amplitudes decrease monotonically, indicating reduced long-range order. The peaks also broaden, reflecting increased lattice disorder, and the peak positions shift closer together in $k$-space, suggesting an increase in average interatomic spacing (reduced density) near the amorphous layer. Multiple CCFs from the amorphous region near the recrystallized layer are shown to illustrate the variability in angular correlations within the phase. Their amplitudes are significantly lower than in the crystalline regions and 2-fold symmetry no longer dominates as strongly. The presence of numerous small peaks suggests local short-range ordering, multiple orientations, and possible residual symmetry fragments within the amorphous matrix.
}
\end{figure*}

\begin{figure*}[!ht]
\centering
\includegraphics[width=0.98\textwidth]{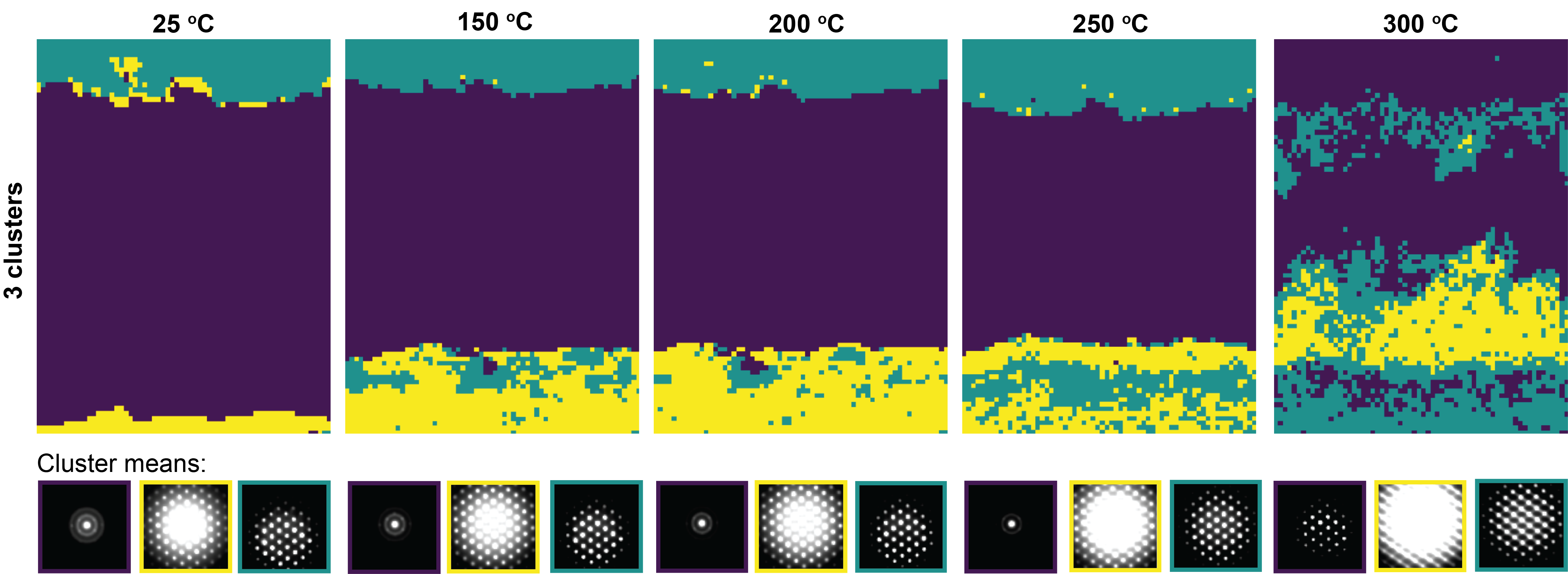}
\caption{\label{fig:figs5} Diffraction patterns from Region 1 were classified using the k-means++ algorithm into two and three clusters. The mean diffraction patterns for each cluster are shown. The classes are computed separately at each temperature step.}
\end{figure*}

%\subsection*{\label{sec:level4} Supplementary Information References}